\documentclass[aps,reprint,amsmath,amssymb,onecolumn,superscriptaddress]{revtex4-2}
\usepackage{graphicx}
\usepackage{dcolumn}
\usepackage{bm}
\usepackage{subfigure}
\usepackage{float}
\usepackage{color}
\usepackage{mathrsfs}
\usepackage{amstext}
\usepackage{booktabs}
\usepackage{siunitx}
\usepackage{lineno}

\begin{document}


\title{Privacy-Enhanced Distributed Sensing Network via Causal-Order Switching and Probe Squeezing }


\author{Zhaotong Cui}
\affiliation{State Key Laboratory of Photonics and Communications, Institute for Quantum Sensing and Information Processing, School of Automation and Intelligent Sensing, Shanghai Jiao Tong University, Shanghai 200240, China}
\author{Binke Xia}
\email{binke97@hku.hk}
\affiliation{QICI Quantum Information and Computation Initiative, School of Computing and Data Science, The University of Hong Kong, Pokfulam Road, Hong Kong, China}
\author{Yuxiang Yang}
\affiliation{QICI Quantum Information and Computation Initiative, School of Computing and Data Science, The University of Hong Kong, Pokfulam Road, Hong Kong, China}
\author{Hongjing Li}
\affiliation{State Key Laboratory of Photonics and Communications, Institute for Quantum Sensing and Information Processing, School of Automation and Intelligent Sensing, Shanghai Jiao Tong University, Shanghai 200240, China}
\affiliation{Hefei National Laboratory, Hefei 230088, China}
\affiliation{Shanghai Research Center for Quantum Sciences, Shanghai 201315, China}
\author{Jingzheng Huang}
\email{jzhuang1983@sjtu.edu.cn}
\affiliation{State Key Laboratory of Photonics and Communications, Institute for Quantum Sensing and Information Processing, School of Automation and Intelligent Sensing, Shanghai Jiao Tong University, Shanghai 200240, China}
\affiliation{Hefei National Laboratory, Hefei 230088, China}
\affiliation{Shanghai Research Center for Quantum Sciences, Shanghai 201315, China}
\author{Guihua Zeng}
\affiliation{State Key Laboratory of Photonics and Communications, Institute for Quantum Sensing and Information Processing, School of Automation and Intelligent Sensing, Shanghai Jiao Tong University, Shanghai 200240, China}
\affiliation{Hefei National Laboratory, Hefei 230088, China}
\affiliation{Shanghai Research Center for Quantum Sciences, Shanghai 201315, China}
\date{\today}
\begin{abstract}
As sensor networks become more interconnected, sensitive information sharing among multiple users poses potential risks of information leakage, making privacy protection a critical challenge. To address this challenge, we quantify network privacy using the maximum information leakage ratio (MILR) and propose a protocol that enhances privacy through causal-order switching and probe squeezing. Causal-order switching harnesses intrinsic network propagation dynamics for privacy encoding, while probe squeezing suppresses information leakage through quantum channels and increases the information available for estimating the global parameter. These mechanisms enable an asymptotic exponential enhancement on the privacy with the squeezing rate. We experimentally demonstrate this mechanism in distributed beam-tilt sensing using spatially structured light without requiring nonclassical light sources. Our work identifies intrinsic network propagation dynamics as a resource for privacy protection and opens a new avenue for continuous-variable sensing networks that combine high precision with enhanced privacy.

\end{abstract}


\maketitle
\section{Introduction}
Spatially distributed sensor networks support sensing and coordination tasks such as clock synchronization~\cite{elson2002fine,schenato2011average}, spatial field estimation~\cite{cortes2009distributed,guestrin2004distributed}, and distributed optical phase estimation~\cite{guo2024distributed}. In classical schemes, each sensor independently estimates its local parameters and transmits the results over a communication channel to a central server, which combines the local estimates to infer a global parameter~\cite{luo2005universal,yan2021distributed}. In cooperative localization and distributed radar networks, for example, sharing local estimates can disclose sensitive details about sensor locations and local observations. Such disclosures may enable targeted attacks that compromise network privacy and sensing reliability~\cite{ristic2023distributed,chen2025analysis}. Sensing networks that handle sensitive information must therefore protect the privacy of local parameters while maintaining the precision of the global estimate.

Without additional privacy safeguards, transmitting these estimates over communication channels may disclose more about individual local parameters than the global sensing task requires~\cite{guo2017estimation,yan2023distributed}. This risk motivates private distributed quantum sensing (DQS)~\cite{PrDQS1}, which aims to estimate a prescribed global function with high precision while restricting access to information about non-target combinations of local parameters.

Recent theoretical studies~\cite{PrDQS2,PrDQS3,PrDQS4,PrDQS5,PrDQS6,PrDQS7} have shown that using nonclassical light  enables privacy-preserving estimation of global functions in sensing networks with a star topology. As networks become more interconnected and their topologies more complex, privacy-preserving protocols must account for how each node encodes its local parameter and how probes propagate through the communication channels along a given route. This motivates the joint design of local parameter encoding and network dynamics to harness probe propagation and routing as resources for controlling information accessibility across the sensing network.

In contrast to the parallel sensing architectures commonly employed in conventional DQS protocols~\cite{DQS1,DQS2,DQS3,DQS4,DQS5,DQS6}, we consider a cyclic network in which a probe prepared at the server sequentially interrogates spatially separated sensors through quantum channels before returning to the server for measurement~\cite{xia2026scaling}. As the probe traverses potentially untrusted channels, its intermediate states may reveal information about non-target combinations of the local parameters. Such leakage corresponds to sensitivity along parameter directions that leave the target quantity locally unchanged. To quantify this leakage relative to the information available about the target global parameter, we introduce the maximum information leakage ratio (MILR), defined as the ratio of the maximum intermediate-state quantum Fisher information (QFI)~\cite{liu2020quantum,helstrom1969quantum,demkowicz2020multi} over normalized non-target parameter directions to the effective QFI for the normalized target global parameter at the server.

To enhance network privacy, we exploit inter-node intrinsic propagation as a dynamical resource for parameter encoding. The noncommutativity between propagation and local sensing generates order-dependent parameter responses, while correlations between the target and nuisance parameters obscure the additional target information in intermediate channel states~\cite{suzuki2020quantum,suzuki2020nuisance}. By combining the complementary information provided by opposite query orders, causal-order switching enables the server to recover this propagation-induced contribution. This recovery can be achieved using a classical mixture of the two orders~\cite{wechs2021quantum,ICO1,ICO2} and requires no multipartite entanglement among the sensing nodes. Probe squeezing further enhances the recovered target global information at the server while suppressing non-target information accessible from intermediate channel states. 

This mechanism enables a quantum-inspired structured-light implementation using coherent optical probes, whose transverse position and momentum uncertainties can be controlled~\cite{wang2024structured}. In the paraxial regime, coherent light and quantum probes share the same mathematical description of spatial-mode evolution and projection measurements. We experimentally demonstrate the protocol for distributed beam-tilt sensing in a free-space optical network by encoding each local tilt as a transverse momentum kick on spatially squeezed first-order Hermite--Gaussian (HG) modes~\cite{PhysRevLett.102.173601,Xia_2023}, with potential applications in interferometric alignment~\cite{gitton2004iris,dooley2013angular} and beam-pointing stabilization~\cite{chapman2025laser}. The noncommutativity between momentum kick encoding and free-space propagation induces an additional transverse displacement of the probe. When the probe sequentially interrogates the sensors along two opposite causal routes selected through polarization control, the average momentum kick is encoded in a route-dependent displacement of its final state. However, this propagation-induced target information remains obscured by correlations between the target and nuisance parameters and cannot be independently extracted.

Our theoretical analysis predicts that, for a fixed average inter-node distance and fixed probe resources, the MILR scales asymptotically as $e^{-4\tau}$ with increasing squeezing rate $\tau$ and decreases further as the number of sensing nodes $M$ increases. Our experiments with $M=2, 3,$ and $4$ sensing nodes and squeezing rate ranging from $\tau=0$ to $\tau=1$ show that the MILR decreases as either the squeezing rate or the network size increases, in agreement with theoretical predictions. These results establish causal-order switching and probe squeezing as complementary resources for privacy-enhanced sensing network, enabling global parameter estimation with reduced relative leakage of non-target information from intermediate states. Without requiring multipartite entanglement among sensing nodes, the protocol offers a practical route toward scalable sensing networks with enhanced privacy.

\section{Results}
\subsection{Theoretical Framework}
Consider an initial probe state $\hat{\rho}_{0}$ that undergoes a unitary transformation $\hat{U}_{\theta}$ encoding the unknown parameter $\theta$, yielding the state $\hat{\rho}_{\theta}=\hat{U}_{\theta}\hat{\rho}_{0}\hat{U}_{\theta}^{\dagger}$. The parameter is subsequently inferred from measurements on the encoded probe. For $n$ independent repetitions, the variance of any locally unbiased estimator $\hat{\theta}$ satisfies the quantum Cramér--Rao bound (QCRB), $\operatorname{Var}(\hat{\theta})\geq [n\mathcal{Q}_{\theta}]^{-1}$, where $\mathcal{Q}_{\theta}$ denotes the QFI of $\hat{\rho}_{\theta}$ with respect to $\theta$. Under standard regularity conditions, this bound can be approached asymptotically using a suitable measurement and an efficient estimator.

In a general DQS task, $M$ individual parameters $\boldsymbol{\theta} = (\theta_1,\theta_2,\ldots,\theta_M)$ are encoded into a quantum probe distributed over a sensing network comprising $M$ spatially separated sensors. The goal of DQS is typically not to estimate all parameters independently, but rather to infer a global function $f(\boldsymbol{\theta})=\bm{\alpha}^{\mathrm{T}}\!\cdot\!\boldsymbol{\theta}$, where $\bm{\alpha}\in\mathbb{R}^{M}$ and $\lVert\bm{\alpha}\rVert_{2}=1$. Throughout this work, we consider a network of $M$ identical sensors, each characterized by the same local sensing Hamiltonian $\hat{G}_{S}$. At the $j$-th sensor, the unknown local parameter $\theta_j$ is encoded onto the probe state through the unitary encoding $\hat{U}_{\theta_j}=\exp(-\mathrm{i}\theta_j\hat{G}_{S})$.

\begin{figure}[htb]
	\centering
	\includegraphics[width=\linewidth]{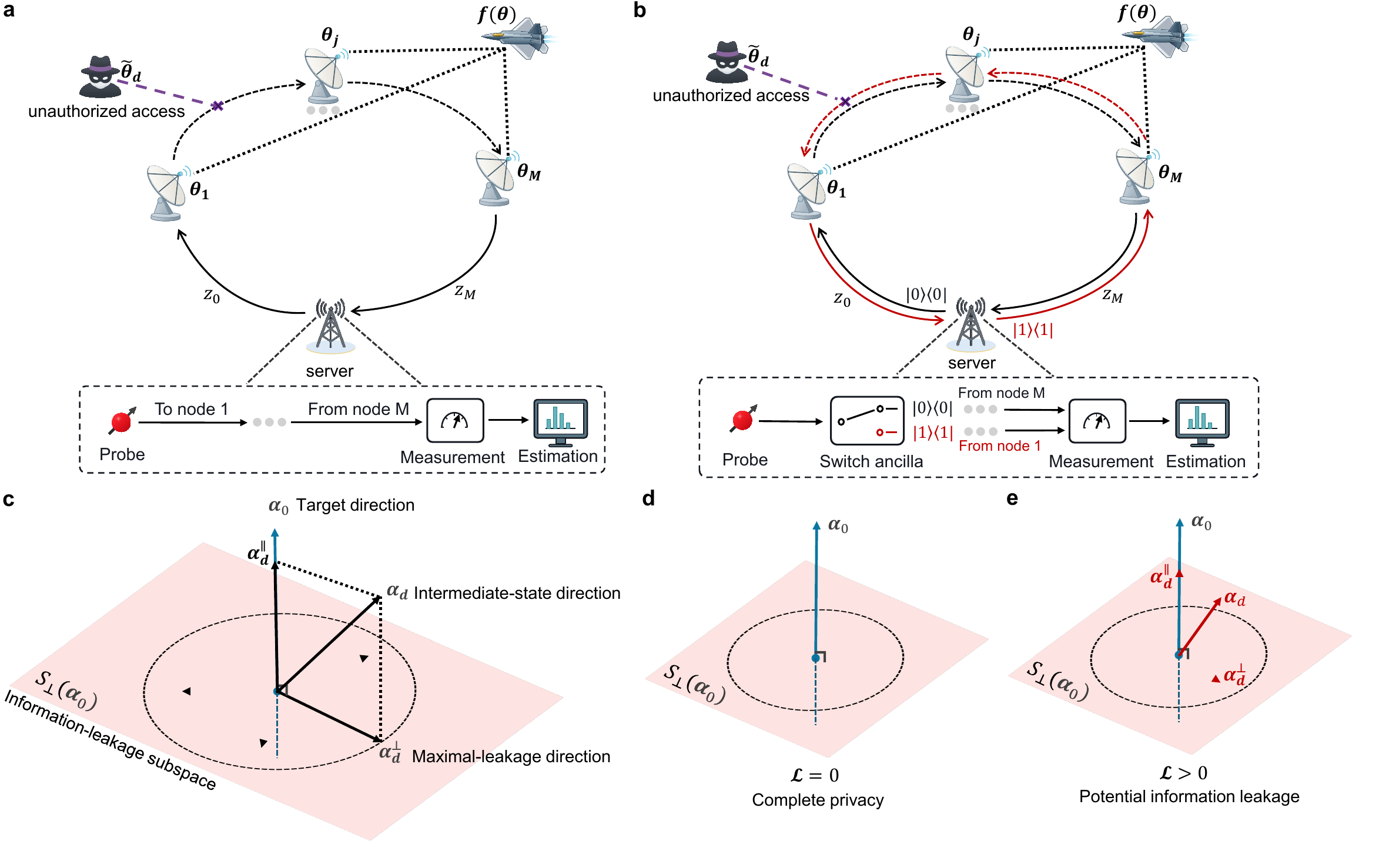}
	\caption{\label{fig:1}Network architectures and geometric interpretation of privacy in DQS. \textbf{a,} Single-order cyclic sensing network. \textbf{b,} Cyclic sensing network with causal-order switching. A control ancilla enables coherent or classical routing, with $|0\rangle$ and $|1\rangle$ labeling the forward (black) and backward (red) query orders, respectively. The server prepares the probe and performs the final measurement. \textbf{c,} Decomposition of the intermediate-state direction $\bm{\alpha}_d$ into components parallel and orthogonal to the target direction $\bm{\alpha}_0$. The orthogonal component $\bm{\alpha}_d^\perp$ identifies the maximal-leakage direction in $\mathcal{S}_{\perp}(\bm{\alpha}_0)$ (pink plane). \textbf{d,} Complete privacy ($\mathcal{L}=0$), with no accessible information along non-target directions at any intermediate stage. \textbf{e,} Potential information leakage ($\mathcal{L}>0$), with nonzero sensitivity to a non-target direction at an accessible intermediate stage.}
\end{figure}

Existing DQS schemes typically consider star-type networks, in which the probe state is distributed to the sensors in parallel~\cite{zhang2021distributed}. Here, we consider a cyclic network comprising a server node and $M$ sensing nodes~\cite{xia2026scaling}, as illustrated in Fig.~\ref{fig:1}a. A probe initially prepared in $|\psi_i\rangle$ at the server node propagates through the inter-node quantum channels and sequentially interrogates the $M$ sensors before returning to the server, where its final state $|\psi_f\rangle$ is measured to estimate the global parameter $f(\boldsymbol{\theta})$. As the probe propagates through potentially untrusted channels, its intermediate states may carry information about non-target combinations of the local parameters. Private DQS aims to estimate a prescribed global function $f(\boldsymbol{\theta})$ with high precision while suppressing information about non-target linear combinations of the local parameters, thereby limiting the precision with which those combinations can be estimated. Assessing privacy in this cyclic network setting therefore requires characterizing potential leakage throughout propagation relative to the useful global information available for the intended sensing task. We therefore introduce the maximum information leakage ratio (MILR), defined as the maximum QFI over all intermediate probe states and all unit-norm parameter directions orthogonal to the target direction, divided by the effective QFI for the normalized target global parameter at the server. The maximization captures the largest QFI associated with any unit-norm parameter direction orthogonal to the target direction across all accessible intermediate probe states.

As illustrated in Fig.~\ref{fig:1}c, we characterize non-target information leakage through the sensitivity of intermediate probe states to parameter variations orthogonal to the target direction. For the target function $f(\bm{\theta})=\bm{\alpha}^{\mathsf{T}}\bm{\theta}$, we define the unit target direction $\bm{\alpha}_0=\bm{\alpha}/\lVert\bm{\alpha}\rVert_2$ and the normalized target parameter $\bar{\theta}_0=\bm{\alpha}_0^{\mathsf{T}}\bm{\theta}$. To quantify the largest potential leakage relative to the effective target information at the server, we define the MILR as
\begin{equation}
\mathcal{L}
=
\frac{
\displaystyle \max_{d}\,
Q\!\left(\tilde{\theta}_{d}^{\perp}\middle|\hat{\rho}_{d}\right)
}{
Q\!\left(\bar{\theta}_{0}\middle|\hat{\rho}_{f}\right)
},
\label{eq:1}
\end{equation}
where $\hat{\rho}_d$ and $\hat{\rho}_f$ denote the intermediate probe state at stage $d$ and the final probe state at the server, respectively. For each intermediate state $\hat{\rho}_d$, the maximum directional leakage QFI is given by
\[
Q\!\left(\tilde{\theta}_{d}^{\perp}\middle|\hat{\rho}_{d}\right)
=
\max_{\substack{
\bm{v}\in\mathcal{S}_\perp(\bm{\alpha}_0)\\
\lVert\bm{v}\rVert_2=1
}}
\bm{v}^{\mathsf{T}}\mathbf{Q}_d\bm{v},
\]
where $\mathbf{Q}_d=Q\!\left(\bm{\theta}\middle|\hat{\rho}_d\right)$ denotes the QFIM of the intermediate probe state $\hat{\rho}_d(\bm{\theta})$ with respect to the local parameter vector $\bm{\theta}$, and $\mathcal{S}_\perp(\bm{\alpha}_0)$ is the subspace orthogonal to the target direction. At each stage $d$, a unit vector attaining this maximum is denoted by $\bm{\alpha}_d^\perp$, which defines the corresponding non-target parameter combination $\tilde{\theta}_d^\perp=(\bm{\alpha}_d^\perp)^{\mathsf{T}}\bm{\theta}$.

As schematically illustrated in Fig.~\ref{fig:1}d, $\mathcal{L}=0$ corresponds to vanishing potential information leakage along every direction in $\mathcal{S}_{\perp}(\bm{\alpha}_0)$ at all accessible intermediate stages. Potential information leakage along any direction in this subspace at an accessible intermediate stage implies $\mathcal{L}>0$, as illustrated in Fig.~\ref{fig:1}e. A smaller MILR indicates lower maximum potential information leakage relative to the effective global information available at the server, corresponding to a higher degree of privacy in the sensing network.

In realistic sensing networks, probe propagation through channels between successive sensing nodes must be incorporated into the overall sensing dynamics. We describe the propagation process via a unitary operator $\hat{U}_{z}=\exp[-\mathrm{i}z\hat{G}_{P}]$ (or $\hat{U}_{t}=\exp[-\mathrm{i}t\hat{G}_{P}]$), where $z$ (or $t$) denotes the propagation distance (or time), and $\hat{G}_{P}$ is the Hermitian generator of the evolution. When $[\hat{G}_{P},\hat{G}_{S}]\neq 0$, inter-node propagation modifies the effective local sensing generators according to the propagation accumulated before each encoding, making the probe’s response to local parameters dependent on the sensor interrogation order. When only a single query order is used, unknown non-target combinations of local parameters enter the propagation-induced probe response as nuisance parameters~\cite{suzuki2020quantum}, whose correlations with the target parameter obscure part of the additional target information. To recover this information, we introduce the causal-order switching scheme illustrated in Fig.~\ref{fig:1}b. The propagation-induced target information remains obscured by nuisance correlations in intermediate channel states but becomes accessible at the server by combining information from opposite query orders. Moreover, by devising an appropriate probe state to alter the uncertainties of the sensing and propagation generators $\hat{G}_{S}$ and $\hat{G}_{P}$, information leakage into the subspace orthogonal to the target global direction can be further suppressed, while the attainable information for estimating the global parameter is simultaneously improved.

Specifically, we first consider the single-order cyclic sensing network illustrated in Fig.~\ref{fig:1}a, in which a probe sequentially interrogates $M$ sensor nodes before returning to the server for measurement. The overall evolution can be described by the unitary operator $\hat{U}_{+}=\hat{U}_{z_{M}}\hat{U}_{\theta_{M}}\hat{U}_{z_{M-1}}\hat{U}_{\theta_{M-1}}\cdots\hat{U}_{\theta_{1}}\hat{U}_{z_{0}}$. We choose the average of the local parameters as the global quantity of interest, with the normalized target direction $\boldsymbol{\alpha}_0=(1,\ldots,1)^{\mathsf{T}}/\sqrt{M}$ and the associated target global parameter $\bar{\theta}_0=\boldsymbol{\alpha}_0^{\mathsf{T}}\boldsymbol{\theta}=\sqrt{M}\bar{\theta}$. At node $j$, the local parameter $\theta_j$ is encoded through $\hat{U}_{\theta_j}=\exp(-\mathrm{i}\theta_j\hat{X})$, which imparts a transverse-momentum kick to the probe. Free-space propagation between nodes is described by $\hat{U}_{z_j}=\exp(-\mathrm{i}z_j\hat{P}^{2}/2k)$, where $z_j$ is the propagation distance and $k$ is the wave number. Such transverse-momentum encoding provides a model for beam-deflection sensing, with applications in precision optical alignment, beam pointing, and angular target localization in optical radar. We denote the propagation distances along the cyclic route by $z_{0},z_{1},\ldots,z_{M}$, where $z_{0}$ connects the server node to the first sensor and $z_{M}$ connects the last sensor back to the server. In the practical sensing network, the total propagation length, $z_{\mathrm{tot}}=\sum_{j=0}^{M}z_j$, and the number of sensor nodes $M$ are generally known in advance. Hence, even if the individual link lengths $z_j$ are not publicly known, their average $\bar{z}$ is public information, as it is completely determined by $z_{\mathrm{tot}}$ and $M$.

For this fixed-order configuration, the MILR is given by $\mathcal{L}_{\mathrm{FCO}}=\frac{\lfloor M^2/4\rfloor}{M^2}$ (see Supplementary Information for details). This value is independent of the initial probe state and therefore remains unchanged under probe squeezing. To exploit the network propagation dynamics for privacy enhancement, we introduce causal-order switching between two opposite interrogation orders, as illustrated in Fig.~\ref{fig:1}b. A control ancilla prepared in the classically mixed state $\hat{\rho}_{\mathrm{c}}^{(\mathrm{SW})}=(|0\rangle\langle 0|+|1\rangle\langle 1|)/2$ selects the forward and backward interrogation orders with equal probability. The backward evolution is described by $\hat{U}_{-}=\hat{U}_{z_{0}}\hat{U}_{\theta_{1}}\hat{U}_{z_{1}}\hat{U}_{\theta_{2}}\cdots\hat{U}_{\theta_{M}}\hat{U}_{z_{M}}$. The joint evolution of the probe and control ancilla is then described by
\begin{equation}
\begin{aligned}
\hat{U}^{(\mathrm{SW})}
&=
\hat{U}_{+}\otimes|0\rangle\langle 0|
+
\hat{U}_{-}\otimes|1\rangle\langle 1|\\
&=
\exp\!\left[-\mathrm{i}\frac{(M+1)\bar{z}}{2k}\hat{P}^{2}\right]
\exp\!\left[-\mathrm{i}\frac{\Gamma_{+}(\boldsymbol{\theta})}{k}\hat{P}\right]
\exp\!\left(-\mathrm{i}M\bar{\theta}\hat{X}\right)
\otimes|0\rangle\langle 0|\\
&\quad+
\exp\!\left[-\mathrm{i}\frac{(M+1)\bar{z}}{2k}\hat{P}^{2}\right]
\exp\!\left[-\mathrm{i}\frac{\Gamma_{-}(\boldsymbol{\theta})}{k}\hat{P}\right]
\exp\!\left(-\mathrm{i}M\bar{\theta}\hat{X}\right)
\otimes|1\rangle\langle 1|.
\end{aligned}\label{eq:2}
\end{equation}
Where
\[
\Gamma_{+}(\boldsymbol{\theta})=\sum_{j=0}^{M-1} z_j\left(\sum_{l=j+1}^{M}\theta_l\right),\quad
\Gamma_{-}(\boldsymbol{\theta})=\sum_{j=1}^{M} z_j\left(\sum_{l=1}^{j}\theta_l\right),\quad
\Gamma_{+}(\boldsymbol{\theta})+\Gamma_{-}(\boldsymbol{\theta})=(M+1)\bar{z}\sqrt{M}\bar{\theta}_0.
\]
and $\bar{z}=\frac{1}{M+1}\sum_{j=0}^{M} z_j$ denotes the average propagation distance between adjacent sensors. Measurement outcomes from both orders are jointly processed at the server to estimate the target global parameter. Consider the initial quantum state is $\hat{\rho}_i=|\psi_i\rangle\langle\psi_i|
$, the final joint state of the probe and the ancilla can be derived as:
\begin{align}
    \hat{\rho}_{f}^{(\mathrm{SW})} &= \hat{U}^{(\mathrm{SW})}\left(|\psi_{i}\rangle\langle\psi_{i}|\otimes\hat{\rho}^{(\mathrm{SW})}\right)\left[\hat{U}^{(\mathrm{SW})}\right]^{\dagger} \nonumber\\
    &= \frac{1}{2}|\psi_{+}\rangle\langle\psi_{+}|\otimes|0\rangle\langle 0|+\frac{1}{2}|\psi_{-}\rangle\langle\psi_{-}|\otimes|1\rangle\langle 1|\label{eq:3}
\end{align}
where $|\psi_{+}\rangle=\hat{U}_{+}|\psi_{i}\rangle$ is the final probe state for forward propagation, while $|\psi_{-}\rangle=\hat{U}_{-}|\psi_{i}\rangle$ is the corresponding state for backward propagation. In this case, the control ancilla acts merely as a classical label for the probe order routing, where $|0\rangle\langle 0|$ denotes forward propagation and $|1\rangle\langle 1|$ denotes backward propagation. The QFIM for the joint estimation of $\bar{\theta}_0$ and $\Gamma_{+}(\boldsymbol{\theta})$ from the final state $\hat{\rho}_{f}^{(\mathrm{SW})}$ is therefore given by
\begin{equation}
\mathcal{Q}\!\left(\bar{\theta}_0,\Gamma_{+}(\boldsymbol{\theta})\,\middle|\,\hat{\rho}_{f}^{(\mathrm{SW})}\right)
=[\mathcal{Q}(\bar{\theta}_0,\Gamma_{+}(\boldsymbol{\theta})||\psi_{+}\rangle)
+\mathcal{Q}(\bar{\theta}_0,\Gamma_{+}(\boldsymbol{\theta})||\psi_{-}\rangle)]/2
\label{eq:4}
\end{equation}
This yields the following partial QFI for estimating $\bar{\theta}_0$ (see Supplementary Information for details):
\begin{equation}
\mathcal{Q}\!\left(\bar{\theta}_0\middle|\hat{\rho}_{f}^{(\mathrm{SW})}\right)
=\frac{M(M+1)^2\bar{z}^{2}(\Delta\hat{P})_{i}^{2}}{k^{2}}
+\frac{4M(M+1)\bar{z}}{k}\operatorname{Cov}_{i}(\hat{X},\hat{P})
+4M(\Delta\hat{X})_{i}^{2},
\label{eq:5}
\end{equation}
where $(\Delta\hat{A})_{i}^{2}$ denotes the variance of operator $\hat{A}$, and $\operatorname{Cov}_{i}(\hat{A},\hat{B})$ denotes the covariance of $\hat{A}$ and $\hat{B}$, both evaluated with respect to the initial probe state $|\psi_{i}\rangle$.

Having quantified the information about the target global parameter accessible at the server, we next evaluate information leakage from intermediate probe states during transmission in the channel. We consider an unauthorized observer accessing the probe after it has interrogated sensors $1,\ldots,d$ in the forward query order. The corresponding intermediate probe state is
\begin{equation}
\hat{\rho}_d(\bm{\theta})=\hat{U}_{d}|\psi_i\rangle\langle\psi_i|
\hat{U}_{d}^{\dagger}=|\psi_d(\bm{\theta})\rangle\langle\psi_d(\bm{\theta})|,
\label{eq:6}
\end{equation}
where $\hat{U}_{d}=\hat{U}_{\theta_{d}}\hat{U}_{z_{d-1}}\hat{U}_{\theta_{d-1}}\cdots\hat{U}_{\theta_{1}}\hat{U}_{z_{0}}$ denotes the probe evolution up to observation stage $d$ under the single query order, and $|\psi_d(\bm{\theta})\rangle=\hat{U}_{d}|\psi_i\rangle$. We characterize the information encoded in the intermediate state $\hat{\rho}_d(\bm{\theta})$ by its QFIM $\mathcal Q\!\left(\bm{\theta}\middle|\hat{\rho}_d\right)$ with respect to the local parameter vector $\bm{\theta}$. The accumulated encoding parameter is $\theta_d=\sum_{j=1}^{d}\theta_j=\boldsymbol{\alpha}_d^{\mathsf T}\bm{\theta}$, where $\boldsymbol{\alpha}_d=(\underbrace{1,\ldots,1}_{d},\underbrace{0,\ldots,0}_{M-d})^{\mathsf T}$. Taking the Schur complement of the nuisance-parameter block associated with $\Gamma_d$ in the joint QFIM for $(\theta_d,\Gamma_d)$ yields the effective QFIM for the local parameter vector $\bm{\theta}$ encoded in the intermediate state $\hat{\rho}_d(\bm{\theta})$ (see Supplementary Information for details):
\begin{equation}
\mathcal{Q}^{(\mathrm{eff})}\!\left(\bm{\theta}\middle|\hat{\rho}_d\right)
=4\left[(\Delta\hat X)_i^2-\frac{\operatorname{Cov}_i^2(\hat X,\hat P)}
{(\Delta\hat P)_i^2}
\right]
\boldsymbol{\alpha}_d\boldsymbol{\alpha}_d^{\mathsf T}.
\label{eq:7}
\end{equation}
Since $\mathcal{Q}^{(\mathrm{eff})}\!\left(\bm{\theta}\middle|\hat{\rho}_d\right)$ is proportional to $\boldsymbol{\alpha}_d\boldsymbol{\alpha}_d^{\mathsf T}$, the normalized direction of maximum leakage within the non-target subspace $\mathcal S_\perp(\boldsymbol{\alpha}_0)$ is the normalized projection of $\boldsymbol{\alpha}_d$ onto this subspace,
\begin{equation}
\widetilde{\boldsymbol{\alpha}}_d^\perp
=\sqrt{\frac{M}{d(M-d)}}
\left(\boldsymbol{\alpha}_d-\frac{d}{\sqrt M}\boldsymbol{\alpha}_0\right),
\qquad 1\le d\le M-1.
\label{eq:8}
\end{equation}
The corresponding scalar parameter is $\widetilde{\theta}_d^\perp=(\widetilde{\boldsymbol{\alpha}}_d^\perp)^{\mathsf T}\bm{\theta}$, and the maximum directional leakage QFI at sensing node $d$ is
\begin{equation}
\begin{aligned}
\mathcal Q_{d,\mathrm{leak}}^{\max}
&=\left(\widetilde{\boldsymbol{\alpha}}_d^\perp\right)^{\mathsf T}
\mathcal{Q}^{(\mathrm{eff})}\!\left(\bm{\theta}\middle|\hat{\rho}_d
\right)\widetilde{\boldsymbol{\alpha}}_d^\perp
=
\frac{4d(M-d)}{M}\left[(\Delta\hat X)_i^2-\frac{\operatorname{Cov}_i^2(\hat X,\hat P)}{(\Delta\hat P)_i^2}\right].
\end{aligned}
\label{eq:9}
\end{equation}
Causal-order switching thus creates an information asymmetry in which the server can combine forward and backward evolutions to recover propagation-induced target information that cannot be independently extracted from intermediate states. Consider an initial Gaussian probe centered at the phase-space origin satisfy $\langle\hat{X}\rangle_i=\langle\hat{P}\rangle_i=0$, and $\operatorname{Cov}_i(\hat{X},\hat{P})=0$. Substituting Eqs.~(\ref{eq:5}) and~(\ref{eq:9}) into Eq.~(\ref{eq:1}) then yields
\begin{equation}
\mathcal{L}_{\mathrm{SW}}
=\frac{4\lfloor M^{2}/4\rfloor}{M^{2}}
\left[
4+\left(\frac{(M+1)\bar{z}}{k}\right)^{2}
\frac{(\Delta\hat{P})_{i}^{2}}{(\Delta\hat{X})_{i}^{2}}
\right]^{-1}.
\label{eq:10}
\end{equation}

The factor $4\lfloor M^{2}/4\rfloor/M^{2}$ equals unity for even $M$ and approaches unity from below for odd $M$ as $M$ increases. The dependence of $\mathcal{L}_{\mathrm{SW}}$ on the variance ratio $(\Delta\hat{P})_{i}^{2}/(\Delta\hat{X})_{i}^{2}$ motivates the use of probe squeezing to further enhance the privacy of network. Since the $N$th-order HG mode corresponds to the Fock state $|N\rangle$, we choose the squeezed first-order HG state $|\psi_{1,\tau}\rangle=|\tau_1\rangle=\hat{S}(\tau)|1\rangle$ as the initial probe, whose transverse wave function can be expanded over odd-order HG modes as
\begin{equation}
\psi_1(x,\tau)=\frac{1}{(\cosh r)^{3/2}}\sum_{K=0}^{\infty}\frac{\sqrt{(2K+1)!}}{2^K K!}\left(-e^{\mathrm{i}\theta}\tanh r\right)^K u_{2K+1}(x),\label{eq:11}
\end{equation}
where $\tau=|\tau|e^{\mathrm{i}\theta}$, with $\tau$ and $\theta$ denoting the squeezing rate and phase, respectively. Here, $u_n(x)=\langle x|n\rangle$ is the normalized transverse wave function of the $n$th-order HG mode at the waist plane, given by
\begin{equation}
u_n(x)=\frac{1}{\sqrt{2^n n!}\,(2\pi\sigma_0^2)^{1/4}}H_n\!\left(\frac{x}{\sqrt{2}\sigma_0}\right)\exp\!\left(-\frac{x^2}{4\sigma_0^2}\right),
\label{eq:12}
\end{equation}
where $H_n$ denotes the $n$th-order Hermite polynomial and $\sigma_0^2$ is the variance of the fundamental Gaussian mode. For the squeezing angle of $\theta=0$, the squeezed first-order HG probe has transverse-momentum and transverse-position variances $(\Delta\hat{P})_i^2=3e^{2r}/(4\sigma_0^2)$ and $(\Delta\hat{X})_i^2=3\sigma_0^2e^{-2r}$, respectively. Substituting these expressions into Eq.~(\ref{eq:10}) yields the privacy enhancement that is substantially increased with increasing squeezing rate $|\tau|$:
\begin{equation}
\mathcal{L}_{\mathrm{SW}}^{(\tau)}=\frac{\lfloor M^2/4\rfloor}{M^2}\left[1+\left(\frac{(M+1)\bar{z}}{4k\sigma_0^2}\right)^2 e^{4|\tau|}\right]^{-1}.
\label{eq:13}
\end{equation}

Importantly, the privacy enhancement of our DQS protocol exploits noncommuting sensing and propagation dynamics to achieve route-dependent parameter encoding. Combining opposite routes enables the server to recover additional target information obscured by target–nuisance coupling in intermediate states. We further show that stronger spatial squeezing and more sensing nodes enhance the effective target information relative to the maximum information leakage from intermediate states. The protocol requires neither multipartite entanglement among sensing nodes nor coherence between different routes, enabling a simple, scalable implementation with coherent optical probes.

\subsection{Structured-Light Implementation for Spatially Squeezed Probes}
For a practical demonstration of the privacy enhancement, we implement an optical DQS network combining causal-order switching with spatially squeezed probes. Throughout this work, we take the average of the local parameters $f(\boldsymbol{\theta})=M^{-1}\sum_{j=1}^{M}\theta_j$ as the global parameter of interest, and quantify privacy in terms of the information leaked about parameter combinations lying in the subspace orthogonal to the target global direction. 

In addition, the propagation distances $z_{\mathrm{in}}$ (from the initial probe state at the beam waist to the input of the network) and $z_{\mathrm{out}}$ (from the output of the network to the detection plane) must be taken into account. These introduce additional propagation evolutions, denoted as $\hat{U}_{z_{\mathrm{in}}}$ and $\hat{U}_{z_{\mathrm{out}}}$, acting on the probe state before entering and after exiting the sensing network, respectively. Throughout the experiments, we ensure that the parameters satisfy $\bar{\theta}\ll 1$, $g_1\ll 1$, and $g_2\ll 1$. Under these conditions, the joint evolution of the probe and the ancilla can be expanded to first order as:
\begin{align}
\hat{U}
&=\hat{U}_{z_{\mathrm{out}}}\hat{U}_{+}\hat{U}_{z_{\mathrm{in}}}\otimes |H\rangle\langle H|+\hat{U}_{z_{\mathrm{out}}}\hat{U}_{-}\hat{U}_{z_{\mathrm{in}}}
\otimes |V\rangle\langle V|\nonumber\\
&\approx
\left(\hat{U}_{z_{\mathrm{tot}}}\otimes\hat{I}_{\mathrm{pol}}\right)
\left[
\left(
\hat{I}_{\mathrm{sp}}
-iM\bar{\theta}\hat{X}
-i\frac{\Gamma_{+}(\boldsymbol{\theta})+\Gamma_{-}(\boldsymbol{\theta})+2z_{\mathrm{in}}M\bar{\theta}}{2k}\hat{P}
\right)\otimes\hat{I}_{\mathrm{pol}}
-i\frac{\Gamma_{+}(\boldsymbol{\theta})-\Gamma_{-}(\boldsymbol{\theta})}{2k}\hat{P}\otimes\hat{A}
\right],\label{eq:14}
\end{align}
where $\hat{A}=|H\rangle\langle H|-|V\rangle\langle V|$ is the Pauli operator, $z_{\mathrm{tot}}=z_{\mathrm{in}}+(M+1)\bar{z}+z_{\mathrm{out}}$ denotes the total propagation distance of the probe. Preparing the probe in $|\psi_i\rangle=|\tau_1\rangle$ and post-selecting the ancilla onto $|f\rangle=|+\rangle=(|H\rangle+|V\rangle)/\sqrt{2}$ yields the final probe state:
\begin{equation}
|\psi_{f}\rangle = 
\frac{\langle f|\hat{U}|i\rangle|\psi_{i}\rangle}{\Vert\langle f|\hat{U}|i\rangle|\psi_{i}\rangle\Vert} 
\approx 
\mathcal{N}_{f}\hat{U}_{z_{\mathrm{tot}}}\left[ |\tau_1\rangle -i\sqrt{3}\,\sigma_{0}M\bar{\theta}\,e^{-r}|\psi_{\hat{X}}\rangle -i\,\frac{\sqrt{3}\,e^{r}}{4\sigma_{0}k} \left(\Gamma_{+}(\boldsymbol{\theta})+\Gamma_{-}(\boldsymbol{\theta})+2z_{\mathrm{in}}M\bar{\theta}\right) |\psi_{\hat{P}}\rangle \right],\label{eq:15}
\end{equation}
where $\mathcal{N}_f$ is a normalization factor, and $|\psi_{\hat{P}}\rangle$ and $|\psi_{\hat{X}}\rangle$ are the normalized states obtained by applying $\hat{P}$ and $\hat{X}$, respectively, to the initial probe state. Their explicit forms are given by
\begin{equation}
\begin{aligned}
|\psi_{\hat{P}}\rangle
&= \frac{i e^{-r}}{\sqrt{3}}
\sum_{K=0}^{\infty} c_K\left(\sqrt{2K+2}\,|2K+2\rangle-\sqrt{2K+1}\,|2K\rangle\right),
\\[6pt]
|\psi_{\hat{X}}\rangle
&= \frac{e^{r}}{\sqrt{3}}\sum_{K=0}^{\infty} c_K\left(\sqrt{2K+1}\,|2K\rangle+\sqrt{2K+2}\,|2K+2\rangle
\right).
\end{aligned}
\label{eq:16}
\end{equation}Since both $|\psi_{\hat{P}}\rangle$ and $|\psi_{\hat{X}}\rangle$ are orthogonal to the initial probe state $|\tau_1\rangle$, Gram–Schmidt orthogonalization yields a normalized detection state that is orthogonal to $|\psi_{\hat{X}}\rangle$ while retaining a nonzero overlap with $|\psi_{\hat{P}}\rangle$. Using $\langle\psi_{\hat{X}}|\psi_{\hat{P}}\rangle=i/3$, we obtain $|\psi_{\hat{X}}^{\perp}\rangle=\frac{1}{2\sqrt{2}}\left(3|\psi_{\hat{P}}\rangle-
i|\psi_{\hat{X}}\rangle\right)$, which is orthogonal to both $|\tau_1\rangle$ and $|\psi_{\hat{X}}\rangle$. We incorporate the common propagation operator by defining the states at the detection plane as $|\tau_1(z_{\mathrm{tot}})\rangle = \hat{U}_{z_{\mathrm{tot}}}|\tau_1\rangle$, $|u_{\hat{P}}(z_{\mathrm{tot}})\rangle= \hat{U}_{z_{\mathrm{tot}}}|\psi_{\hat{P}}\rangle$, and $|u_{\hat{X}}(z_{\mathrm{tot}})\rangle =\hat{U}_{z_{\mathrm{tot}}}|\psi_{\hat{X}}\rangle$. In terms of these propagated states, the final probe state can be written as
\begin{equation}
|\psi_f(z_{\mathrm{tot}})\rangle \approx \mathcal{N}_f \Bigg[ |\tau_1(z_{\mathrm{tot}})\rangle - i\sqrt{3}\,\sigma_0 M\bar{\theta}e^{-r}|u_{\hat{X}}(z_{\mathrm{tot}})\rangle - i\frac{\sqrt{3}\,e^{r}}{4\sigma_0 k}\left(\Gamma_{+}(\boldsymbol{\theta})+\Gamma_{-}(\boldsymbol{\theta})+2z_{\mathrm{in}}M\bar{\theta}\right)|u_{\hat{P}}(z_{\mathrm{tot}})\rangle \Bigg].\label{eq:17}
\end{equation}

The normalized detection state at the output plane is given by $|u_{\hat{X}}^{\perp}(z_{\mathrm{tot}})\rangle=\frac{1}{2\sqrt{2}}\left[3|u_{\hat{P}}(z_{\mathrm{tot}})\rangle-i|u_{\hat{X}}(z_{\mathrm{tot}})\rangle\right]$. Owing to the preservation of inner products under unitary propagation, this state remains orthogonal to both $|\tau_1(z_{\mathrm{tot}})\rangle$ and $|u_{\hat{X}}(z_{\mathrm{tot}})\rangle$. The server implements the positive-operator-valued measure (POVM) $\hat{\Pi}^{\perp}(z_{\mathrm{tot}})=\{\hat{\Pi}_1(z_{\mathrm{tot}}),\hat{\Pi}_2(z_{\mathrm{tot}})\}$, where $\hat{\Pi}_1(z_{\mathrm{tot}})=|u_{\hat{X}}^{\perp}(z_{\mathrm{tot}})\rangle\langle u_{\hat{X}}^{\perp}(z_{\mathrm{tot}})|$ and $\hat{\Pi}_2(z_{\mathrm{tot}})=\hat{I}_{\mathrm{sp}}-\hat{\Pi}_1(z_{\mathrm{tot}})$. We express the information associated with the experimental signal $\bar{\theta}$ along the normalized global direction $\bm{\alpha}_0$ using $\bar{\theta}_0=\bm{\alpha}_0^{\mathsf{T}}\bm{\theta}=\sqrt{M}\,\bar{\theta}$. Applying the chain rule to the classical Fisher information (CFI) calculated from the outcome probabilities (see Methods part) gives
\begin{equation}
\begin{aligned}
\mathcal{F}(\bar{\theta}_0|\hat{\rho}_{f})
=\left(\frac{\partial\bar{\theta}}{\partial\bar{\theta}_0}\right)^2
\mathcal{F}(\bar{\theta}|\hat{\rho}_{f})
\simeq\frac{2e^{2r}}{3\sigma_0^2k^2}
M\left[(M+1)\bar{z}+2z_{\mathrm{in}}\right]^2.
\end{aligned}
\label{eq:18}
\end{equation}

To quantify the potential maximum information leakage, we consider the intermediate probe state after the $d$-th sensing interaction, $|\psi_d\rangle=\hat{U}_d\hat{U}_{z_{\mathrm{in}}}|\psi_i\rangle$, we quantify potential leakage by optimizing the effective CFI over all POVMs on the intermediate state $\hat{\rho}_d(\bm{\theta})=|\psi_d(\bm{\theta})\rangle\langle\psi_d(\bm{\theta})|$, where $|\psi_d(\bm{\theta})\rangle=\hat{U}_d\hat{U}_{z_{\mathrm{in}}}|\psi_i\rangle$ is the probe state after the $d$-th sensing interaction. Within the local estimation framework, this optimum is given by the corresponding partial QFI. Maximizing over intermediate stages yields
\begin{equation}
\mathcal{F}_{\mathrm{leak}}^{\max}
= \max_{1\leq d\leq M-1}\sup_{\{\hat{\Pi}_x\}}
\mathcal{F}^{(\mathrm{eff})}\!\left(\tilde{\theta}_d^\perp\middle|\hat{\rho}_d;\{\hat{\Pi}_x\}\right)
= \max_{1\leq d\leq M-1}
\mathcal{Q}\!\left(\tilde{\theta}_d^\perp\middle|\hat{\rho}_d\right)
\simeq 12\sigma_0^2e^{-2r}\frac{\lfloor M^2/4\rfloor}{M}.
\label{eq:19}
\end{equation}
Substituting the target CFI and maximum leakage CFI from Eqs.~\eqref{eq:18} and~\eqref{eq:19} into the Eq.~\eqref{eq:1} yields MILR with the following expression for our experimental scheme:
\begin{equation}
\mathcal{L}
=\frac{\mathcal{F}_{\mathrm{leak}}^{\max}}
{\mathcal{F}_{\mathrm{global}}}
=
\frac{18\sigma_0^4k^2e^{-4r}}
{\left[(M+1)\bar{z}+2z_{\mathrm{in}}\right]^2}
\frac{\lfloor M^2/4\rfloor}{M^2}.
\label{eq:20}
\end{equation}

Thus, probe squeezing enhances the target CFI while suppressing intermediate-state leakage, reducing the MILR as $\mathcal{L}\propto e^{-4r}$. Increasing the number of sensing nodes further reduces the MILR, yielding the asymptotic scaling $\mathcal{L}\propto M^{-2}$ at fixed average inter-node propagation distance $\bar{z}$, input and output propagation distances, and measurement resources.

\subsection{Experimental Setup and Results}
\begin{figure}[htp]
	\centering
	\includegraphics[width=0.95\linewidth]{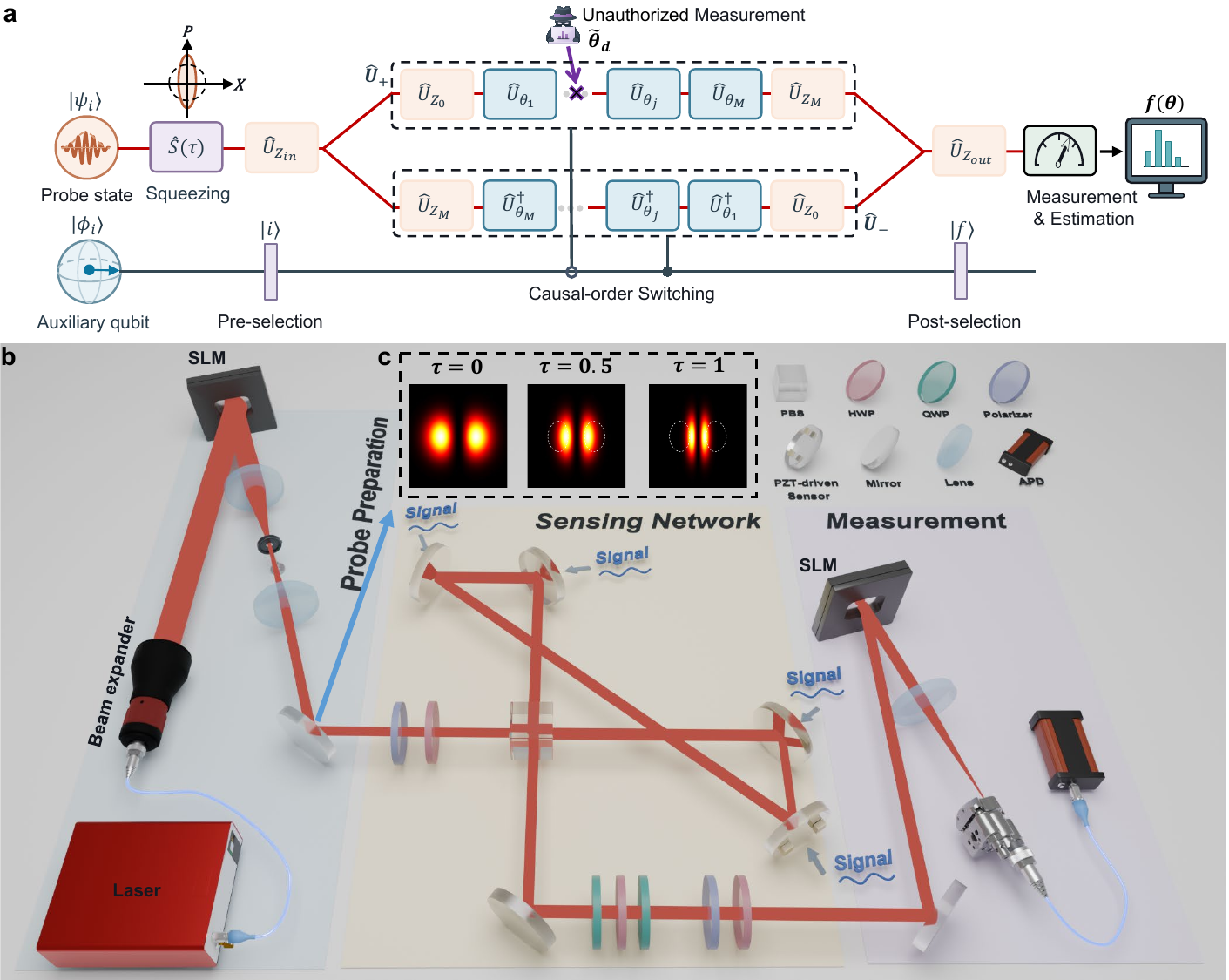}
	\caption{\label{fig:2}Privacy-enhanced distributed beam-tilt sensing protocol and its optical implementation. \textbf{a,} Causal-order switching. An auxiliary qubit preselected in $|i\rangle$ controls the forward and backward query orders, $\hat{U}_{+}$ and $\hat{U}_{-}$, of a spatially squeezed probe. After postselection in $|f\rangle$, the probe is measured to estimate $f(\boldsymbol{\theta})$. The purple arrow marks a potential unauthorized measurement of an intermediate probe state that may disclose non-target information. \textbf{b,} Optical implementation. A spatial light modulator (SLM) and a spatial filter prepare a spatially squeezed first-order HG probe. A polarization Sagnac interferometer routes the probe through four PZT-driven mirrors in opposite orders. A second SLM, a Fourier-transform lens, and a single-mode fiber implement spatial-mode projection, followed by photon detection with an avalanche photodiode. \textbf{c,} Transverse probe intensity profiles at $\tau=0$, $0.5$, and $1$, from left to right.
}
\end{figure}

We implement the proposed sensing protocol in a distributed beam-tilt experiment. As shown in Fig.~\ref{fig:2}a, the transverse spatial mode of the beam serves as the sensing probe, while its polarization controls the order in which the probe interrogates the sensing nodes. We prepare the probe in a spatially squeezed first-order HG mode $|\psi_{1,\tau}\rangle$ and preselect its polarization in the $45^\circ$ linear state $|i\rangle=(|H\rangle+|V\rangle)/\sqrt{2}$. The horizontal and vertical polarization components traverse the same sensing nodes in opposite orders. After the probe returns to the server, we postselect its polarization in $|f\rangle$ and measure its transverse spatial mode to estimate the target function $f(\boldsymbol{\theta})$. The purple arrow indicates where an unauthorized observer could access an intermediate probe state.

The optical implementation uses a triangular polarization Sagnac interferometer with up to four sensing nodes as shown in Fig.~\ref{fig:2}b. The triangular geometry avoids unwanted parity-induced sign reversals in the encoded transverse momentum shifts. At the $j$th sensing node, a piezoelectric transducer (PZT) drives a mirror that deflects the beam by a small angle $\varphi_j$. This deflection imparts a transverse momentum kick that encodes the local parameter $\theta_j=k\varphi_j$.

Spatially squeezed first-order HG probes with different squeezing rate $\tau$ are prepared using a spatial light modulator (SLM) followed by a spatial filtering system. The polarization ancilla is initialized in $|i\rangle$ using a linear polarizer and a half-wave plate (HWP), and postselected onto $|f\rangle$ using a second linear polarizer. A quarter-wave plate (QWP)–HWP–QWP sequence compensates for the unwanted relative phase accumulated between the two polarization-controlled routes. The operating principles and specific wave-plate configurations are detailed in the Supplemental Material. Following postselection, a second SLM, together with a Fourier-transform lens and a single-mode fiber (SMF), implements the projective measurement through spatial mode filtering. Photons coupled into the SMF are detected by an avalanche photodiode (APD).

By displaying the projection $\hat{\Pi}_1(z_{\mathrm{tot}})$ on the SLM, the parameter $\bar{\theta}$ encoded by the evolution can be detected with projection probability:
\begin{equation}
P_{\bar{\theta}}
= \langle \psi_f(z_{\mathrm{tot}}) |
\hat{\Pi}_1(z_{\mathrm{tot}})
| \psi_f(z_{\mathrm{tot}}) \rangle
\simeq \frac{e^{2r}M^{2}\left[(M+1)\bar{z}+2z_{\mathrm{in}}\right]^2}
{6\sigma_{0}^{2}k^{2}}\,\bar{\theta}^{2}.
\label{eq:21}
\end{equation}

The information extracted from the observed measurement outcomes can be characterized by the observed Fisher information (OFI). Let $N_0$ denote the effective number of trials inferred from the reference optical power and $N_1$ the effective number of successful projection events inferred from the DC readout. Within the independent-trial model, the log-likelihood is $\ell(\bar{\theta})=N_1\ln P(\bar{\theta})+(N_0-N_1)\ln[1-P(\bar{\theta})]+C$, where $C$ is independent of $\bar{\theta}$. At a locally identifiable interior maximum-likelihood estimate $\widehat{\bar{\theta}}$, the fitted probability satisfies $P(\widehat{\bar{\theta}})=N_1/N_0$. The OFI is then
\begin{equation}
F_{\mathrm{obs}}\bigl(\widehat{\bar{\theta}}\bigr) = -\left.\frac{\partial^2\ell(\bar{\theta})}{\partial\bar{\theta}^2}\right|_{\bar{\theta}=\widehat{\bar{\theta}}} = N_0\frac{\bigl[P'(\widehat{\bar{\theta}})\bigr]^2}{P(\widehat{\bar{\theta}})\bigl[1-P(\widehat{\bar{\theta}})\bigr]} \simeq \frac{4N_1}{\bigl(\widehat{\bar{\theta}}\bigr)^2}.\label{eq:22}
\end{equation}
The final approximation uses the quadratic response in the weak-encoding regime, $P(\bar{\theta})\propto\bar{\theta}^2$, and assumes $P(\widehat{\bar{\theta}})\ll 1$.

In the experiment, PZT-driven mirrors encode the local sensing parameters through small angular deflections of the probe beam. Synchronized sinusoidal signals at $f=2\,\mathrm{kHz}$ drive the PZTs, producing synchronous angular modulation across the sensing nodes. Static mirror misalignment introduces a non-negligible offset $\bar{\theta}^{\Delta}$ in the target average parameter, such that $\bar{\theta}^{\mathrm{tot}}=\bar{\theta}^{\Delta}+\bar{\theta}\cos(2\pi f t)$, where $\bar{\theta}$ denotes the modulation amplitude. We operate in the regime $|\bar{\theta}|\ll|\bar{\theta}^{\Delta}|$, corresponding to a small modulation around the bias operating point.

Before applying the drive signals, we project the output probe onto $|u_{\tau}(z_{\mathrm{tot}})\rangle$ and record the reference optical power $I_0$ to determine the effective sample number $N_0$ for each different squeezing rate $\tau$. We then apply the drive signals and project the output probe onto $|u_{\hat{X}}^{\perp}(z_{\mathrm{tot}})\rangle$, recording the corresponding detected optical power $I_1$. By varying $\bar{\theta}$ and fitting the measured response, we obtain the magnitude of the static offset $\bar{\theta}^{\Delta}$ from
\begin{equation}
\frac{\partial I_1^{(2\,\mathrm{kHz})}}
{\partial \bar{\theta}}
\simeq
\frac{e^{2r}M^2
\left[(M+1)\bar{z}+2z_{\mathrm{in}}\right]^2}
{3\sigma_0^2k^2}
I_0|\bar{\theta}^{\Delta}|.
\label{eq:23}
\end{equation}
The sinusoidal modulation is used to calibrate the local response of the measurement. The Fisher information reported below characterizes the estimation of a static perturbation about the calibrated bias operating point. The DC component of the post-selected readout provides the signal associated with successful projection events, from which $N_{1}$ is determined. Together with the calibrated bias $\bar{\theta}^{\Delta}$, this allows the local OFI at the bias operating point to be calculated. Under the maximum-likelihood and low-probability conditions specified above, we obtain
\begin{equation}
\mathcal{F}_{\mathrm{obs}}\!\left(\bar{\theta}_0\right)
=
\frac{1}{M}
\mathcal{F}_{\mathrm{obs}}\!\left(\bar{\theta}^{\mathrm{tot}}\right)
\simeq
\frac{4N_{1}}
{M\left(\bar{\theta}^{\Delta}\right)^2}.
\label{eq:24}
\end{equation}

\begin{figure}[hbt]
	\centering
	\includegraphics[width=0.85\linewidth]{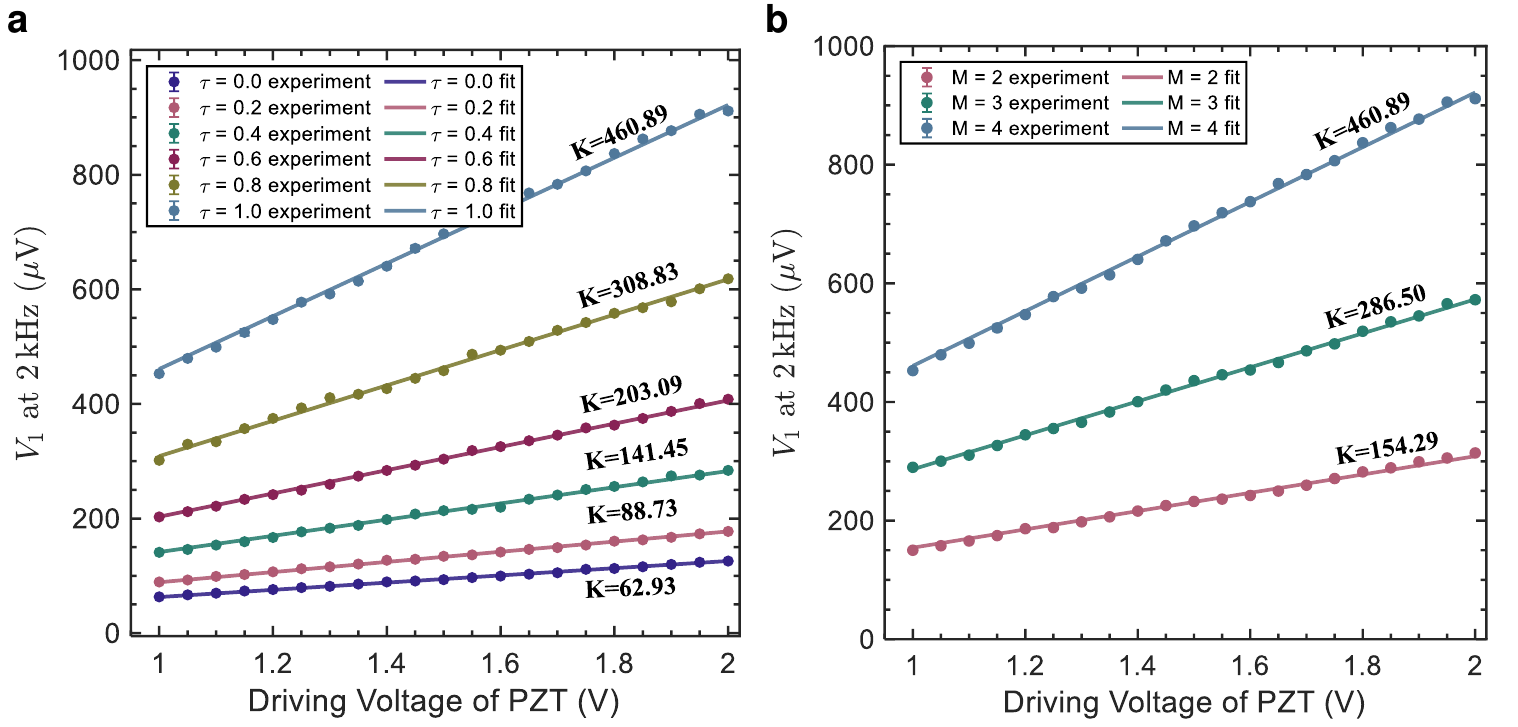}
	\caption{\label{fig:3}Measured response signals and linear fits for different squeezing rate $\tau$ and network sizes $M$ from the spectrum analyzer. The applied peak-to-peak voltage is varied from \SI{1}{\volt} to \SI{2}{\volt} in \SI{50}{\milli\volt} increments, corresponding to an average beam-tilt modulation $\bar{\varphi}$ ranging from \SI{2.2}{\micro\radian} to \SI{4.4}{\micro\radian} in \SI{0.11}{\micro\radian} increments. \textbf{\textsf{a}}, Detected signal levels of different squeezing rate $\tau$ at a fixed network size $M=4$. \textbf{\textsf{b}}, Detected signal levels for different network sizes $M=2,3,4$ at a fixed squeezing rate $\tau=1$. Circular markers with error bars indicate the experimental data, and solid lines show linear fits.}
\end{figure}

Fig.~\ref{fig:3} shows the measured responses and corresponding fits for varying squeezing rate $\tau$ at $M=4$, and for varying network sizes $M$ at $\tau=1$. By linearly fitting the measured $V_1^{(2\,\mathrm{kHz})}$ as a function of the parameter $\bar{\theta}$, we extracted the response slope $K$ and determined the magnitude of the static offset $\bar{\theta}^{\Delta}$ using Eq.~\eqref{eq:23} for each network configuration.

In our experiment, the reference optical power $I_0$ incident on the APD was fixed at approximately \SI{566}{\pico\watt}, corresponding to a photon flux of approximately $2.22\times10^{9}$ photons/s at a wavelength of \SI{780}{\nano\meter}. We then apply the drive signals and project the output probe onto $|u_{\hat{X}}^{\perp}(z_{\mathrm{tot}})\rangle$, recording the corresponding detected optical power $I_1$. A peak-to-peak PZT drive voltage of \SI{1}{\volt} produced a beam-tilt modulation of \SI{2.2}{\micro\radian}, corresponding to an encoded parameter $\theta = k\varphi = 17.72\,\mathrm{m}^{-1}$ (see the Methods section for details). The APD converted the detected optical power into an electrical voltage signal, from which a spectrum analyzer extracted the component at \SI{2}{\kilo\hertz}, corresponding to the optical-power modulation $V_1^{(2\,\mathrm{kHz})}=\gamma I_1^{(2\,\mathrm{kHz})}$, where $\gamma$ is the conversion gain, which is the product of the responsivity and transresistance gain of the APD. We synchronously varied the peak-to-peak drive voltages applied to the PZTs from \SI{1}{\volt} to \SI{2}{\volt} in \SI{50}{\milli\volt} increments, corresponding to an average beam-tilt modulation $\bar{\varphi}$ ranging from \SI{2.2}{\micro\radian} to \SI{4.4}{\micro\radian} in \SI{0.11}{\micro\radian} increments.

To assess the information accessible from intermediate probe states, we first consider a spatial-mode projection that removes the leading-order nuisance contribution associated with $\hat{P}$. The corresponding normalized detection mode is $|\psi_{\hat{P}}^{\perp}\rangle = \frac{1}{2\sqrt{2}}\left(3|\psi_{\hat{X}}\rangle + i|\psi_{\hat{P}}\rangle\right)$, which satisfies $\langle\psi_{\hat{P}}^{\perp}|\psi_{\hat{P}}\rangle = 0$. For this projection, $\left|\langle\psi_{\hat{P}}^{\perp}|\psi_{\hat{X}}\rangle\right|^2 = 8/9$, so only $8/9$ of the leading-order $\hat{X}$-encoded signal probability is retained. Within the weak-encoding model, the resulting CFI therefore falls below the corresponding effective QFI with nuisance parameters taken into account. To avoid underestimating the potential leakage due to this restricted readout, we evaluate it using the effective QFI of the intermediate probe states. Maximizing this quantity over normalized parameter directions orthogonal to the target direction and over intermediate positions $d$, and using the same effective sample number $N_0$ as for the server measurement, we obtain
\begin{equation}
    \mathcal{F}_{\mathrm{leak}}^{\max}
    = N_0 \max_{1 \leq d \leq M-1}
      \mathcal{F}\!\left(
          \tilde{\boldsymbol{\theta}}_{d}^{\perp}
          \,\middle|\, \hat{\rho}_{d}
      \right)
    = 12 N_0 \sigma_0^2 e^{-2r}
      \frac{\lfloor M^2/4 \rfloor}{M}.
    \label{eq:25}
\end{equation}

\begin{figure}[b]
	\centering
	\includegraphics[width=\linewidth]{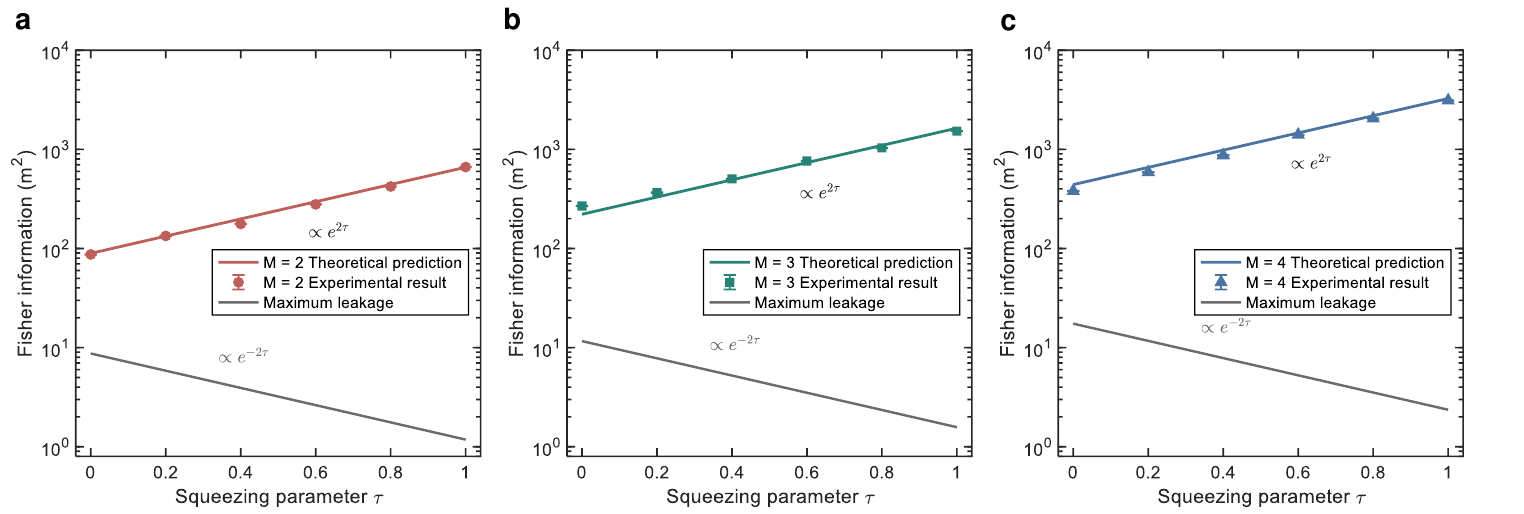}
\caption{\label{fig:4}
Experimental result of information leakage and global-parameter information.
\textbf{a-c} Specific experimental results for the global parameter and maximum
information leakage for $M=2$, $3$, and $4$, respectively. Colored solid lines represent the theoretical Fisher information for the global parameter. Red circles, green squares, and blue triangles with error bars indicate the experimental results for $M=2$, $3$, and $4$, respectively, and gray solid lines denote the maximum leakage Fisher information calculated using the same measured input photon number $N_0$.
}
\end{figure}

Fig.~\ref{fig:4}\textbf{\textsf{a--c}} compares the target global-parameter Fisher information with the maximum intermediate-state leakage for networks containing $M=2$, $3$, and $4$ sensing nodes, respectively. The measured target information agrees with the theoretical predictions, while the leakage estimates are obtained from the intermediate-state model using the measured input photon number $N_0$. For fixed $M$ and $N_0$, the target-parameter FI scales as $e^{2\tau}$, whereas the maximum leakage FI scales as $e^{-2\tau}$. Probe squeezing thus increases the information available for global-parameter estimation while reducing the estimated information accessible from intermediate probe states. Moreover, at fixed average inter-node separation and equal photon resources, the target information grows more rapidly with $M$ than the maximum leakage, further reducing leakage relative to useful target information. These contrasting dependencies explain the predicted scaling $\mathcal{L}_{\mathrm{exp}}\propto e^{-4\tau}$ and its additional reduction with increasing network size, which are examined separately below. The results provide a quantitative basis for assessing the sensing precision and privacy enhancement within the considered intermediate-state access model.

Combining Eq.~(\ref{eq:24}) and~(\ref{eq:25}), the experimentally evaluated MILR can be calculated as
\begin{equation}
\mathcal{L}_{\mathrm{exp}}
=
\frac{\mathcal{F}_{\mathrm{leak}}^{\max}}
{\mathcal{F}_{\mathrm{obs}}(\bar{\theta}_0)}
\simeq
3N_0\sigma_0^2
\bigl(\bar{\theta}^{\Delta}\bigr)^{2}
\frac{\lfloor M^2/4\rfloor e^{-2\tau}}{N_1}.
\label{eq:26}
\end{equation}

\begin{figure}[hbt]
	\centering
	\includegraphics[width=0.8\linewidth]{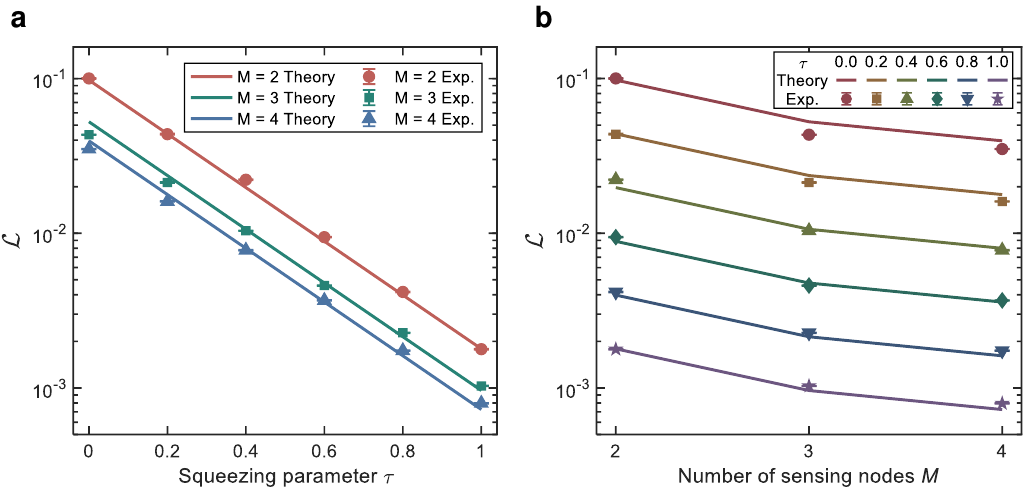}
\caption{\label{fig:5}
Dependence of the MILR on probe squeezing and network size.
\textbf{a}, Experimental MILR, $\mathcal{L}_{\mathrm{exp}}$, as a function of the squeezing rate $\tau$ for networks with $M=2$, $3$, and $4$ sensing nodes. For each network, the ratio decreases with increasing squeezing, following the predicted exponential scaling $\mathcal{L}_{\mathrm{exp}}\propto e^{-4\tau}$. \textbf{b}, Dependence of $\mathcal{L}_{\mathrm{exp}}$ on $M$ at fixed squeezing parameters $\tau=0,0.2,\ldots,1.0$. At fixed average inter-node separation, increasing the number of sensing nodes from two to four further reduces the ratio at every squeezing setting.
Solid lines show the theoretical predictions, while filled markers denote ratios calculated from the maximum intermediate-state leakage FI and the experimentally measured global-parameter FI. Error bars represent standard errors propagated from the fitted probability derivatives.
}
\end{figure}

Fig.~\ref{fig:5} summarizes the dependence of the experimentally evaluated MILR, $\mathcal{L}$, on spatial squeezing and network size. Fig.~\ref{fig:5}\textbf{\textsf{a}} shows that, for each network with $M=2$, $3$, or $4$ sensing nodes, the ratio decreases with increasing squeezing rate $\tau$, following the predicted exponential scaling $\mathcal{L}\propto e^{-4\tau}$. Fig.~\ref{fig:5}\textbf{\textsf{b}} presents the complementary dependence on $M$ at six fixed squeezing settings, revealing a further reduction as the number of sensing nodes increases from two to four. At fixed average inter-node separation, this reduction arises because the target-parameter information grows more rapidly with $M$ than the maximum intermediate-state leakage. Both trends are consistent with Eq.~\eqref{eq:26}, supporting the predicted reduction of leakage relative to target information through spatial squeezing and increased network size within the considered intermediate-state access model.

\section{Discussion}
Our results demonstrate that causal-order switching and probe squeezing can enhance privacy in sensing networks by controlling the accessibility of information throughout the network. The noncommutativity between local sensing and propagation through inter-node channels generates route-dependent parameter encoding. Combining the forward and backward query orders modifies the correlation structure between the target and nuisance parameters, enabling the server to access propagation-induced target information that is obscured by nuisance parameters in a single order route. This mechanism requires only a classical mixture of opposite causal orders, without multipartite entanglement among the sensing nodes or coherence between the routes. By increasing the probe's momentum variance and reducing its position variance, squeezing further enhances the target information while suppressing potential information leakage within the sensing network.

We take the local momentum displacements introduced at individual sensing nodes as the sensing parameters and implement a polarization-controlled Sagnac sensing network using spatially squeezed first-order HG probes. Our theoretical analysis shows that the MILR decreases with increasing spatial squeezing rate $\tau$ and exhibits an asymptotic scaling of $e^{-4\tau}$ when the squeezing-dependent contribution dominates. For the spatial-mode readout employed here~\cite{10.1063/1.4869819}, the ratio of leakage information to target readout information decreases as $e^{-4\tau}$ within the perturbative approximation, with experimental results consistent with this prediction. Furthermore, at a fixed average distance between adjacent sensing nodes, theory predicts that this ratio decreases with increasing node number and asymptotically scales as $M^{-2}$ in the large-$M$ limit. These findings indicate that probe squeezing and network size can jointly improve privacy in distributed sensing.

The decrease of the MILR with increasing node number $M$ assumes a fixed average distance between adjacent nodes and therefore accompanies an increase in the spatial extent and total propagation length of the network. Such growth is intrinsic to network expansion under this assumption. Conventional private DQS models~\cite{PrDQS1,PrDQS2} often treat transmission through quantum channels primarily as a means of probe distribution, without explicitly exploiting its contribution to parameter encoding. The key distinction therefore lies not in whether propagation length grows with network size, but in whether propagation is actively converted into accessible sensing information through noncommuting dynamics. Although our experiment considers momentum displacements in the transverse spatial degree of freedom of an optical beam, the theoretical framework, formulated in terms of canonical variables and their noncommutativity, also applies to displacement estimation with coherent light in optical phase space. Our theoretical and experimental results therefore contribute to the development and refinement of estimation theory for privacy enhancement in DQS and suggest potential applications in distributed radio-frequency electric-field measurements~\cite{PhysRevLett.124.150502} and optomechanical force sensing~\cite{DQS6}, offering an approach to estimating global signals while suppressing leakage of local parameter information.

\section{Methods}
\subsection{Experimental materials}
In our experiments, we employed a single-frequency laser (NKT Photonics, Koheras HARMONIK series) operating at $\SI{780}{\nm}$. Spatially squeezed HG beams were generated using a spatial light modulator (SLM; Hamamatsu Photonics, model X13138-02) with $1272 \times 1024$ pixels and a pixel pitch of $\SI{12.5}{\micro\metre}$. The Fourier lens in the $4f$ system had a focal length of $\SI{10}{\cm}$. A circular pinhole with a diameter of $\SI{200}{\micro\metre}$ served as the spatial filter. A Glan–Taylor polarizer (GTP) from Thorlabs and an HWP from LBTEK are used to initialize the polarization state of light as $\ang{45}$-polarized state. We set up a polarized triangular Sagnac interferometer to implement the cyclic sensing network with a coherent superposition of opposite causal orders, where the number of sensors in the network can be scaled by adjusting the interferometer. (See the Supplemental Materials for details.) The number of sensing nodes could be varied by reconfiguring the interferometer. Two QWPs and an HWP, all from LBTEK, compensated for the unwanted relative phase induced by the interferometer. (Details of the network configuration and phase compensation are provided in the Supplemental Material.)

To generate an angular tilt signal on each sensing node, we pasted 2 PZT chips along the horizontal plane with an interval of $\SI{20}{\mm}$ on the back of a mirror. (See the Supplemental Materials for detailed design of sensor.) These PZT chips are from Core Tomorrow Company (Part No. NAC2013). A driving voltage of $1\,\mathrm{V}_{\mathrm{pp}}$ produced a mirror tilt of $\SI{1.1}{\micro\radian}$, resulting in an angular deflection of the reflected beam of $\varphi = \SI{2.2}{\micro\radian}$ and a corresponding transverse momentum kick of $k\varphi = \SI{17.72}{\per\metre}$, where $k = 2\pi/\lambda = \SI{8.06e6}{\per\metre}$ is the wave number of the laser beam. The two PZT chips on each signal mirror were driven by sinusoidal signals at $f=\SI{2}{\kHz}$ with a relative phase difference of $\pi$, generated using the waveform generator of Moku:Pro from Liquid Instruments. In this experiment, we modulated the waist radius of the fundamental Gaussian beam as $w_0 = 2\sigma_0 = \SI{160}{\micro\metre}$. Residual misalignment of the PZT-actuated mirrors introduces an unavoidable static angular offset $\bar{\theta}^{\Delta}$ in our experiment. This offset must therefore be included in the measurement model when evaluating the Fisher information.

In practice, by projecting the final beam state onto $|u_{\tau}(z_{\mathrm{tot}})\rangle$, we determined the effective number of trials, $N_0=\dot{N}T_{\mathrm{det}}$, from the reference optical power measured by projecting the output beam onto $|u_\tau(z_{\mathrm{tot}})\rangle$. Here, $\dot{N}$ is the photon flux and $T_{\mathrm{det}}$ is the effective detection time. We detected the projection signal with a silicon avalanche photodetector (APD410A, Thorlabs) with a maximum conversion gain of $2.65\times10^7\,\mathrm{V/W}$ and a bandwidth of $10\,\mathrm{MHz}$. The reference optical power was fixed at $I_0=566\,\mathrm{pW}$ across the different squeezing rate, corresponding to a photon flux of $\dot{N}=2.22\times10^9\,\mathrm{s}^{-1}$ at $780\,\mathrm{nm}$. We analyzed the detected voltage using the spectrum analyzer module of Moku:Pro (Liquid Instruments). The resolution bandwidth of $9.78\,\mathrm{Hz}$ corresponded to $T_{\mathrm{det}}=102.25\,\mathrm{ms}$, giving $N_0=2.27\times10^8$.

\subsection{Nuisance parameter estimation and partial QFI}

In this work, we formulate the estimation of the prescribed global function $f(\bm{\theta})$ as a single global function estimation task with unknown nuisance parameters, rather than as a general multiparameter estimation problem. More precisely, the original parameter vector \(\bm{\theta}\) can be reparameterized as $\bm{\eta}=(\eta_{1},\eta_{2},\dots,\eta_{M})^{\mathsf{T}}$, The first coordinate, \(\eta_{\mathrm{I}}\equiv\eta_{1}=f(\bm{\theta})\), is identified as the parameter of interest, whereas the remaining coordinates, \(\bm{\eta}_{\mathrm{N}}=(\eta_{2},\ldots,\eta_{M})^{\mathsf{T}}\), provide a complementary parametrization of the remaining degrees of freedom and are treated as nuisance parameters. Statistical correlations between \(\eta_{\mathrm{I}}\) and \(\bm{\eta}_{\mathrm{N}}\) may reduce the attainable precision in estimating the global function. In this setting, the QFIM \(\mathcal{Q}(\bm{\theta})\) and \(\mathcal{Q}(\bm{\eta})\) contain equivalent local statistical information under the reparameterization, the relevant QCRB cannot, in general, be inferred from the single-parameter QFI associated with \(f(\bm{\theta})\) alone. Instead, the influence of the nuisance parameters must be explicitly accounted for through the full reparameterized QFIM. The resulting effective information for estimating \(\eta_{\mathrm{I}}\) is quantified by the partial QFI, equivalently given by the Schur complement of the nuisance-parameter block.
\begin{equation}
    \mathcal{Q}_{\eta_{\mathrm{I}}|\bm{\eta}_{\mathrm{N}}}(\bm{\eta}) = \mathcal{Q}_{\eta_{\mathrm{I}};\eta_{\mathrm{I}}}(\bm{\eta})-\mathcal{Q}_{\eta_{\mathrm{I}};\bm{\eta}_{\mathrm{N}}}(\bm{\eta})\left[\mathcal{Q}_{\bm{\eta}_{\mathrm{N}};\bm{\eta}_{\mathrm{N}}}(\bm{\eta})\right]^{-1}\mathcal{Q}_{\bm{\eta}_{\mathrm{N}};\eta_{\mathrm{I}}}(\bm{\eta}), 
\end{equation}
For $n$ independently prepared probes, an estimator that is locally unbiased for $\eta_{\mathrm{I}}$ with respect to the full parameter vector satisfies
\begin{equation}
\operatorname{Var}(\hat{\eta}_{\mathrm{I}})
\geq
\frac{1}{
n\,\mathcal{Q}_{\eta_{\mathrm{I}}|\bm{\eta}_{\mathrm{N}}}(\bm{\eta})
}.
\end{equation}
For a single scalar parameter of interest, this bound is locally tight under measurement optimization and can be approached asymptotically using suitable adaptive measurements and efficient estimators under standard regularity conditions.

\subsection{FI calculation}

For a normalized output state $|\psi_{\boldsymbol{g}}\rangle$ parameterized by $\boldsymbol{g}$, a measurement described by the POVM $\{\hat{\Pi}_{\lambda}(z_{\mathrm{tot}})\}_{\lambda}$ yields outcome $\lambda$ with probability $p_{\lambda}(\boldsymbol{g})=\langle\psi_{\boldsymbol{g}}|\hat{\Pi}_{\lambda}(z_{\mathrm{tot}})|\psi_{\boldsymbol{g}}\rangle$. The associated classical Fisher information matrix (CFIM) has elements
\begin{equation}
\mathcal{F}_{ij}=
\sum_{\lambda}
\frac{1}{p_{\lambda}(\boldsymbol{g})}
\frac{\partial p_{\lambda}(\boldsymbol{g})}{\partial g_i}
\frac{\partial p_{\lambda}(\boldsymbol{g})}{\partial g_j}.
\label{eq:20}
\end{equation}

When the parameter vector depends on a scalar parameter $\xi$ through a differentiable mapping $\boldsymbol{g}=\boldsymbol{g}(\xi)$, the corresponding scalar CFI follows from the chain rule,
\begin{equation}
\mathcal{F}(\xi)=\sum_{\lambda}\frac{1}{p_{\lambda}}\left(\frac{\partial p_{\lambda}}{\partial\xi}\right)^2=\sum_{i,j}\frac{\partial g_i}{\partial\xi}\mathcal{F}_{ij}(\boldsymbol{g})\frac{\partial g_j}{\partial\xi}=\left(\frac{\partial\boldsymbol{g}}{\partial\xi}\right)^{\mathsf{T}}\mathcal{F}(\boldsymbol{g})\left(\frac{\partial\boldsymbol{g}}{\partial\xi}\right).
\end{equation}
Here, $p_{\lambda}=p_{\lambda}[\boldsymbol{g}(\xi)]$, and $\partial\boldsymbol{g}/\partial\xi$ denotes the column vector with components $\partial g_i/\partial\xi$.

For the experimental evaluation, the measurement response is calibrated using the measured optical powers and modulation-induced signal slopes. The reference and projected optical powers determine the effective trial number $N_0$ and successful-event number $N_1$, respectively. Within the independent-trial projection model, the corresponding log-likelihood for the parameter $\theta$ is
\begin{equation}
\ell(\theta)
=
N_1\ln P(\theta)
+
(N_0-N_1)\ln[1-P(\theta)]
+C,
\end{equation}
where $\theta$ denotes the total average parameter in the local static-estimation model, and the calibration parameters are held fixed. The observed Fisher information (OFI) is defined by the negative curvature of this log-likelihood at the maximum-likelihood estimate,
\begin{equation}
\mathcal{F}_{\mathrm{obs}}(\hat{\theta})
=
-\left.
\frac{\partial^2\ell(\theta)}{\partial\theta^2}
\right|_{\theta=\hat{\theta}}.
\end{equation}
At a locally identifiable interior maximum-likelihood point satisfying $P(\hat{\theta})=N_1/N_0$, this becomes
\begin{equation}
\mathcal{F}_{\mathrm{obs}}(\hat{\theta})
=
N_0
\frac{[P'(\hat{\theta})]^2}
{P(\hat{\theta})[1-P(\hat{\theta})]}.
\end{equation}
Thus, for this model, the OFI equals the total CFI evaluated at the fitted parameter value. The theoretical curves use the predicted measurement probabilities, whereas the experimental evaluation uses the calibrated response and photon-equivalent event numbers inferred from the measured optical powers.

\begin{acknowledgments}
	This work was supported by the National Natural Science Foundation of China (Grant No. 62471289), the Excellent Young Scientists Fund (Hong Kong and Macau) of the National Natural Science Foundation of China (Project No. 12322516), the Guangdong Provincial Quantum Science Strategic Initiative (Project No. GDZX2605003), and the Research Grants Council of Hong Kong (RGC) through its General Research Fund (GRF; grant 17302724).
\end{acknowledgments}

\noindent {\bf Data and Materials Availability} All data needed to evaluate the conclusions in the paper are present in the paper and the Supplementary Materials.

\noindent {\bf Competing Interests} All authors declare that they have no competing interests.

\noindent {\bf Author Contribution} Z.C. and B.X. designed the scheme. J.H. and G.Z. supervised the research project. Z.C., B.X., and Y.Y. constructed the theoretical model. Z.C. and J.H. carried out the experiments with assistance from H.L. Z.C. analyzed the experimental data. Z.C. wrote the manuscript with assistance from the other authors. All authors have read and approved the final version of the manuscript.

\noindent {\bf Corresponding Authors:} Binke Xia, Jingzheng Huang

\bibliography{bibliography}

@article{DQS1,
  title = {Multiparameter Estimation in Networked Quantum Sensors},
  author = {Proctor, Timothy J. and Knott, Paul A. and Dunningham, Jacob A.},
  journal = {Phys. Rev. Lett.},
  volume = {120},
  issue = {8},
  pages = {080501},
  numpages = {6},
  year = {2018},
  month = {Feb},
  publisher = {American Physical Society},
  doi = {10.1103/PhysRevLett.120.080501},
  url = {https://link.aps.org/doi/10.1103/PhysRevLett.120.080501}
}

@article{DQS2,
   author = {Gessner, Manuel and Smerzi, Augusto and Pezzè, Luca},
   title = {Multiparameter squeezing for optimal quantum enhancements in sensor networks},
   journal = {Nature Communications},
   volume = {11},
   number = {1},
   pages = {3817},
   ISSN = {2041-1723},
   DOI = {10.1038/s41467-020-17471-3},
   url = {https://doi.org/10.1038/s41467-020-17471-3},
   year = {2020},
   type = {Journal Article}
}

@article{DQS3,
   author = {Liu, Li-Zheng and Zhang, Yu-Zhe and Li, Zheng-Da and Zhang, Rui and Yin, Xu-Fei and Fei, Yue-Yang and Li, Li and Liu, Nai-Le and Xu, Feihu and Chen, Yu-Ao and Pan, Jian-Wei},
   title = {Distributed quantum phase estimation with entangled photons},
   journal = {Nature Photonics},
   volume = {15},
   number = {2},
   pages = {137-142},
   ISSN = {1749-4893},
   DOI = {10.1038/s41566-020-00718-2},
   url = {https://doi.org/10.1038/s41566-020-00718-2},
   year = {2021},
   type = {Journal Article}
}

@article{DQS4,
   author = {Kim, Dong-Hyun and Hong, Seongjin and Kim, Yong-Su and Kim, Yosep and Lee, Seung-Woo and Pooser, Raphael C. and Oh, Kyunghwan and Lee, Su-Yong and Lee, Changhyoup and Lim, Hyang-Tag},
   title = {Distributed quantum sensing of multiple phases with fewer photons},
   journal = {Nature Communications},
   volume = {15},
   number = {1},
   pages = {266},
   ISSN = {2041-1723},
   DOI = {10.1038/s41467-023-44204-z},
   url = {https://doi.org/10.1038/s41467-023-44204-z},
   year = {2024},
   type = {Journal Article}
}

@article{DQS5,
   author = {Guo, Xueshi and Breum, Casper R. and Borregaard, Johannes and Izumi, Shuro and Larsen, Mikkel V. and Gehring, Tobias and Christandl, Matthias and Neergaard-Nielsen, Jonas S. and Andersen, Ulrik L.},
   title = {Distributed quantum sensing in a continuous-variable entangled network},
   journal = {Nature Physics},
   volume = {16},
   number = {3},
   pages = {281-284},
   ISSN = {1745-2481},
   DOI = {10.1038/s41567-019-0743-x},
   url = {https://doi.org/10.1038/s41567-019-0743-x},
   year = {2020},
   type = {Journal Article}
}

@article{DQS6,
   author = {Xia, Yi and Agrawal, Aman R. and Pluchar, Christian M. and Brady, Anthony J. and Liu, Zhen and Zhuang, Quntao and Wilson, Dalziel J. and Zhang, Zheshen},
   title = {Entanglement-enhanced optomechanical sensing},
   journal = {Nature Photonics},
   volume = {17},
   number = {6},
   pages = {470-477},
   ISSN = {1749-4893},
   DOI = {10.1038/s41566-023-01178-0},
   url = {https://doi.org/10.1038/s41566-023-01178-0},
   year = {2023},
   type = {Journal Article}
}

@article{PrDQS1,
  title={Private network parameter estimation with quantum sensors},
  author={Shettell, Nathan and Hassani, Majid and Markham, Damian},
  journal={arXiv preprint arXiv:2207.14450},
  year={2022}
}

@article{PrDQS2,
  title={Privacy in networks of quantum sensors},
  author={Hassani, Majid and Scheiner, Santiago and Paris, Matteo GA and Markham, Damian},
  journal={Physical Review Letters},
  volume={134},
  number={3},
  pages={030802},
  year={2025},
  publisher={APS}
}

@article{PrDQS3,
  title={Anonymous and private parameter estimation in networks of quantum sensors},
  author={de Jong, Jarn and Scheiner, Santiago and Solomons, Naomi R and Chaoui, Ziad and Markham, Damian and Pappa, Anna},
  journal={Physical Review Applied},
  volume={24},
  number={5},
  pages={054053},
  year={2025},
  publisher={APS}
}

@article{PrDQS4,
  author  = {Alushi, Uesli and Di Candia, Roberto},
  title   = {Privacy in distributed quantum sensing with {Gaussian} quantum networks},
  journal = {npj Quantum Information},
  year    = {2026},
  volume  = {12},
  number  = {1},
  pages   = {132},
  doi     = {10.1038/s41534-026-01266-3}
}

@article{PrDQS5,
  title={Privacy in continuous-variable distributed quantum sensing},
  author={de Oliveira Junior, A and Andersen, Anton L and Larsen, Benjamin Lundgren and Moore, Sean William and Markham, Damian and Takeoka, Masahiro and Brask, Jonatan Bohr and Andersen, Ulrik L},
  journal={PRX Quantum},
  volume={7},
  number={3},
  pages={033036},
  year={2026},
  publisher={APS}
}

@article{PrDQS6,
  title={Quantum-private distributed sensing},
  author={Ho, Joseph and Webb, Jonathan W and Brooks, Russell MJ and Grasselli, Federico and Gauger, Erik and Fedrizzi, Alessandro},
  journal={Journal of Physics: Photonics},
  volume={8},
  number={2},
  pages={025006},
  year={2026},
  publisher={IOP Publishing}
}

@article{PrDQS7,
  title={Universal operational privacy in distributed quantum sensing},
  author={Namkung, Min and Kim, Dong-Hyun and Hong, Seongjin and Kim, Yong-Su and Lee, Su-Yong and Lim, Hyang-Tag},
  journal={Physical Review Applied},
  volume={26},
  number={1},
  pages={014044},
  year={2026},
  publisher={APS}
}

@article{ICO1,
  title = {Quantum Metrology with Indefinite Causal Order},
  author = {Zhao, Xiaobin and Yang, Yuxiang and Chiribella, Giulio},
  journal = {Phys. Rev. Lett.},
  volume = {124},
  issue = {19},
  pages = {190503},
  numpages = {6},
  year = {2020},
  month = {May},
  publisher = {American Physical Society},
  doi = {10.1103/PhysRevLett.124.190503},
  url = {https://link.aps.org/doi/10.1103/PhysRevLett.124.190503}
}

@article{ICO2,
  author  = {Yin, Peng and Zhao, Xiaobin and Yang, Yuxiang and Guo, Yu
             and Zhang, Wen-Hao and Li, Gong-Chu and Han, Yong-Jian
             and Liu, Bi-Heng and Xu, Jin-Shi and Chiribella, Giulio
             and Chen, Geng and Li, Chuan-Feng and Guo, Guang-Can},
  title   = {Experimental super-{Heisenberg} quantum metrology with indefinite gate order},
  journal = {Nature Physics},
  year    = {2023},
  volume  = {19},
  number  = {8},
  pages   = {1122--1127},
  doi     = {10.1038/s41567-023-02046-y}
}

@article{xia2026scaling,
  title={Scaling enhancement in distributed quantum sensing via causal order switching},
  author={Xia, Binke and Cui, Zhaotong and Huang, Jingzheng and Yang, Yuxiang and Zeng, Guihua},
  journal={arXiv preprint arXiv:2601.14708},
  year={2026}
}

@article{PhysRevLett.102.173601,
  title = {Ultrasensitive Beam Deflection Measurement via Interferometric Weak Value Amplification},
  author = {Dixon, P. Ben and Starling, David J. and Jordan, Andrew N. and Howell, John C.},
  journal = {Phys. Rev. Lett.},
  volume = {102},
  issue = {17},
  pages = {173601},
  numpages = {4},
  year = {2009},
  month = {Apr},
  publisher = {American Physical Society},
  doi = {10.1103/PhysRevLett.102.173601},
  url = {https://link.aps.org/doi/10.1103/PhysRevLett.102.173601}
}

@article{Xia_2023,
   author = {Xia, Binke and Huang, Jingzheng and Li, Hongjing and Wang, Han and Zeng, Guihua},
   title = {Toward incompatible quantum limits on multiparameter estimation},
   journal = {Nature Communications},
   volume = {14},
   number = {1},
   pages = {1021},
   ISSN = {2041-1723},
   DOI = {10.1038/s41467-023-36661-3},
   url = {https://doi.org/10.1038/s41467-023-36661-3},
   year = {2023},
   type = {Journal Article}
}

@article{wang2024structured,
  title={Structured light analogy of quantum squeezed states},
  author={Wang, Zhaoyang and Zhan, Ziyu and Vetlugin, Anton N and Ou, Jun-Yu and Liu, Qiang and Shen, Yijie and Fu, Xing},
  journal={Light: Science \& Applications},
  volume={13},
  number={1},
  pages={297},
  year={2024},
  publisher={Nature Publishing Group UK London}
}

@article{wechs2021quantum,
  title={Quantum circuits with classical versus quantum control of causal order},
  author={Wechs, Julian and Dourdent, Hippolyte and Abbott, Alastair A and Branciard, Cyril},
  journal={PRX Quantum},
  volume={2},
  number={3},
  pages={030335},
  year={2021},
  publisher={APS}
}

@article{PhysRevLett.124.150502,
  title = {Demonstration of a Reconfigurable Entangled Radio-Frequency Photonic Sensor Network},
  author = {Xia, Yi and Li, Wei and Clark, William and Hart, Darlene and Zhuang, Quntao and Zhang, Zheshen},
  journal = {Phys. Rev. Lett.},
  volume = {124},
  issue = {15},
  pages = {150502},
  numpages = {6},
  year = {2020},
  month = {Apr},
  publisher = {American Physical Society},
  doi = {10.1103/PhysRevLett.124.150502},
  url = {https://link.aps.org/doi/10.1103/PhysRevLett.124.150502}
}

@article{10.1063/1.4869819,
    author = {Sun, Hengxin and Liu, Kui and Liu, Zunlong and Guo, Pengliang and Zhang, Junxiang and Gao, Jiangrui},
    title = {Small-displacement measurements using high-order Hermite-Gauss modes},
    journal = {Applied Physics Letters},
    volume = {104},
    number = {12},
    pages = {121908},
    year = {2014},
    month = {03},
    issn = {0003-6951},
    doi = {10.1063/1.4869819},
    url = {https://doi.org/10.1063/1.4869819}
}

@article{yan2021distributed,
  title={Distributed confidentiality fusion estimation against eavesdroppers},
  author={Yan, Xinhao and Zhang, Yuchen and Xu, Daxing and Chen, Bo},
  journal={IEEE Transactions on Aerospace and Electronic Systems},
  volume={58},
  number={4},
  pages={3633--3642},
  year={2021},
  publisher={IEEE}
}

@article{guo2017estimation,
  title={Estimation in wireless sensor networks with security constraints},
  author={Guo, Xiaoxi and Leong, Alex S and Dey, Subhrakanti},
  journal={IEEE Transactions on Aerospace and Electronic Systems},
  volume={53},
  number={2},
  pages={544--561},
  year={2017},
  publisher={IEEE}
}

@article{yan2023distributed,
  title={Distributed encryption fusion estimation against full eavesdropping},
  author={Yan, Xinhao and Chen, Bo and Zhang, Yuchen and Yu, Li},
  journal={Automatica},
  volume={153},
  pages={111025},
  year={2023},
  publisher={Elsevier}
}

@article{liu2020quantum,
  title={Quantum Fisher information matrix and multiparameter estimation},
  author={Liu, Jing and Yuan, Haidong and Lu, Xiao-Ming and Wang, Xiaoguang},
  journal={Journal of Physics A: Mathematical and Theoretical},
  volume={53},
  number={2},
  pages={023001},
  year={2020},
  publisher={IOP Publishing}
}

@article{helstrom1969quantum,
  title={Quantum detection and estimation theory},
  author={Helstrom, Carl W},
  journal={Journal of statistical physics},
  volume={1},
  number={2},
  pages={231--252},
  year={1969},
  publisher={Springer}
}

@article{demkowicz2020multi,
  title={Multi-parameter estimation beyond quantum Fisher information},
  author={Demkowicz-Dobrza{\'n}ski, Rafa{\l} and G{\'o}recki, Wojciech and Gu{\c{t}}{\u{a}}, M{\u{a}}d{\u{a}}lin},
  journal={Journal of Physics A: Mathematical and Theoretical},
  volume={53},
  number={36},
  pages={363001},
  year={2020},
  publisher={IOP Publishing}
}

@article{suzuki2020quantum,
  title={Quantum state estimation with nuisance parameters},
  author={Suzuki, Jun and Yang, Yuxiang and Hayashi, Masahito},
  journal={Journal of Physics A: Mathematical and Theoretical},
  volume={53},
  number={45},
  pages={453001},
  year={2020},
  publisher={IOP Publishing}
}

@article{suzuki2020nuisance,
  title={Nuisance parameter problem in quantum estimation theory: Tradeoff relation and qubit examples},
  author={Suzuki, Jun},
  journal={Journal of Physics A: Mathematical and Theoretical},
  volume={53},
  number={26},
  pages={264001},
  year={2020},
  publisher={IOP Publishing}
}

@inproceedings{gitton2004iris,
  title={IRIS: an infrared tilt sensor for the VLTI},
  author={Gitton, Philippe B and Leveque, Samuel A and Avila, Gerardo and Duc, Thanh Phan},
  booktitle={Proceedings of SPIE},
  volume={5491},
  pages={944--953},
  year={2004}
}

@article{dooley2013angular,
  title={Angular control of optical cavities in a radiation-pressure-dominated regime: the Enhanced LIGO case},
  author={Dooley, Katherine L and Barsotti, Lisa and Adhikari, Rana X and Evans, Matthew and Fricke, Tobin T and Fritschel, Peter and Frolov, Valera and Kawabe, Keita and Smith-Lefebvre, Nicol{\'a}s},
  journal={Journal of the Optical Society of America A},
  volume={30},
  number={12},
  pages={2618--2626},
  year={2013},
  publisher={Optical Society of America}
}

@article{chapman2025laser,
  author  = {Chapman, L. and Tzoganis, V. and Murray, III, S. and Liu, Y.},
  title   = {Laser beam pointing stabilization using analog position-sensitive diodes},
  journal = {Review of Scientific Instruments},
  year    = {2025},
  volume  = {96},
  number  = {8},
  pages   = {083002},
  doi     = {10.1063/5.0263787}
}

@article{zhang2021distributed,
  title={Distributed quantum sensing},
  author={Zhang, Zheshen and Zhuang, Quntao},
  journal={Quantum Science \& Technology},
  volume={6},
  number={4},
  pages={043001},
  year={2021},
  publisher={IOP Publishing}
}

@article{elson2002fine,
  title={Fine-grained network time synchronization using reference broadcasts},
  author={Elson, Jeremy and Girod, Lewis and Estrin, Deborah},
  journal={ACM SIGOPS Operating Systems Review},
  volume={36},
  number={SI},
  pages={147--163},
  year={2002},
  publisher={ACM, New York, NY, USA}
}

@article{schenato2011average,
  title={Average TimeSynch: A consensus-based protocol for clock synchronization in wireless sensor networks},
  author={Schenato, Luca and Fiorentin, Federico},
  journal={Automatica},
  volume={47},
  number={9},
  pages={1878--1886},
  year={2011},
  publisher={Elsevier}
}

@article{cortes2009distributed,
  title={Distributed Kriged Kalman filter for spatial estimation},
  author={Cort{\'e}s, Jorge},
  journal={IEEE Transactions on Automatic Control},
  volume={54},
  number={12},
  pages={2816--2827},
  year={2009},
  publisher={IEEE}
}

@inproceedings{guestrin2004distributed,
  title={Distributed regression: an efficient framework for modeling sensor network data},
  author={Guestrin, Carlos and Bodik, Peter and Thibaux, Romain and Paskin, Mark and Madden, Samuel},
  booktitle={Proceedings of the 3rd international symposium on Information processing in sensor networks},
  pages={1--10},
  year={2004}
}

@article{guo2024distributed,
  title={Distributed phase estimation at the Heisenberg limit with classical light},
  author={Guo, Zijian and Sun, Peijie and Sun, Yifan and Li, Qian and Kong, Lingjun and Zhang, Xiangdong},
  journal={Advanced Photonics Research},
  volume={5},
  number={3},
  pages={2300223},
  year={2024},
  publisher={Wiley Online Library}
}

@article{ristic2023distributed,
  title={Distributed range-only localization that preserves sensor and navigator privacies},
  author={Ristic, Marko and Noack, Benjamin and Hanebeck, Uwe D},
  journal={IEEE Transactions on Automatic Control},
  volume={68},
  number={12},
  pages={7151--7163},
  year={2023},
  publisher={IEEE}
}

@article{chen2025analysis,
  title={An analysis and simulation of security risks in radar networks from the perspective of cybersecurity},
  author={Chen, Runyang and Zhang, Yi and Li, Xiuhe and Ran, Jinhe},
  journal={Sensors},
  volume={25},
  number={17},
  pages={5239},
  year={2025},
  publisher={MDPI}
}

@article{luo2005universal,
  title={Universal decentralized estimation in a bandwidth constrained sensor network},
  author={Luo, Zhi-Quan},
  journal={IEEE Transactions on information theory},
  volume={51},
  number={6},
  pages={2210--2219},
  year={2005},
  publisher={IEEE}
}

\end{document}



\title{Supplementary Materials for Privacy-Enhanced Distributed Sensing Network via Causal-Order Switching and Probe Squeezing }


\author{Zhaotong Cui}
\affiliation{State Key Laboratory of Photonics and Communications, Institute for Quantum Sensing and Information Processing, School of Automation and Intelligent Sensing, Shanghai Jiao Tong University, Shanghai 200240, China}
\author{Binke Xia}
\email{binke97@hku.hk}
\affiliation{QICI Quantum Information and Computation Initiative, School of Computing and Data Science, The University of Hong Kong, Pokfulam Road, Hong Kong, China}
\affiliation{Hefei National Laboratory, Hefei 230088, China}
\affiliation{Shanghai Research Center for Quantum Sciences, Shanghai 201315, China}
\author{Yuxiang Yang}
\affiliation{QICI Quantum Information and Computation Initiative, School of Computing and Data Science, The University of Hong Kong, Pokfulam Road, Hong Kong, China}
\author{Hongjing Li}
\affiliation{State Key Laboratory of Photonics and Communications, Institute for Quantum Sensing and Information Processing, School of Automation and Intelligent Sensing, Shanghai Jiao Tong University, Shanghai 200240, China}
\affiliation{Hefei National Laboratory, Hefei 230088, China}
\affiliation{Shanghai Research Center for Quantum Sciences, Shanghai 201315, China}
\author{Jingzheng Huang}
\email{jzhuang1983@sjtu.edu.cn}
\affiliation{State Key Laboratory of Photonics and Communications, Institute for Quantum Sensing and Information Processing, School of Automation and Intelligent Sensing, Shanghai Jiao Tong University, Shanghai 200240, China}
\affiliation{Hefei National Laboratory, Hefei 230088, China}
\affiliation{Shanghai Research Center for Quantum Sciences, Shanghai 201315, China}
\author{Guihua Zeng}
\affiliation{State Key Laboratory of Photonics and Communications, Institute for Quantum Sensing and Information Processing, School of Automation and Intelligent Sensing, Shanghai Jiao Tong University, Shanghai 200240, China}
\affiliation{Hefei National Laboratory, Hefei 230088, China}
\affiliation{Shanghai Research Center for Quantum Sciences, Shanghai 201315, China}



\maketitle
\section{Quantum Fisher Information in Cyclic Sensing Network}
\subsection{QFIM and partial QFI}

In a general distributed quantum sensing (DQS) task, $M$ individual parameters $\boldsymbol{\theta} = (\theta_1,\theta_2,\ldots,\theta_M)$ are encoded into a quantum probe distributed over a sensing network comprising $M$ spatially separated sensors. The goal of DQS is typically not to estimate all parameters independently, but rather to infer a global function $f(\boldsymbol{\theta})=\bm{\alpha}^{\mathrm{T}}\!\cdot\!\boldsymbol{\theta}$, where $\bm{\alpha}\in\mathbb{R}^{M}$ and $\lVert\bm{\alpha}\rVert_{2}=1$. Therefore, DQS can be formulated as a single-parameter estimation problem, with the remaining independent combinations of the local parameters treated as nuisance parameters. The objective of privacy-enhanced DQS is to increase the information available for estimating the target global parameter while limiting information leakage about these non-target combinations.

We use the partial quantum Fisher information (QFI), derived from the corresponding quantum Fisher information matrix (QFIM), to quantify the information available for estimating the parameter of interest in the presence of nuisance parameters. Let $\hat{\rho}(\bm{\theta})$ denote a quantum state parameterized by the unknown vector $\bm{\theta}=(\theta_1,\theta_2,\ldots,\theta_M)^{\mathsf{T}}$. For the joint estimation of these parameters, the $(j,k)$ element of the QFIM is given by
\begin{equation}
    \mathcal{Q}_{jk}(\bm{\theta})
    = \frac{1}{2}\mathrm{Tr}\left(\hat{\rho}(\bm{\theta})\,\{\hat{L}_{j},\hat{L}_{k}\}\right),
    \label{eq:S1}
\end{equation}
where $\{\hat{A},\hat{B}\}=\hat{A}\hat{B}+\hat{B}\hat{A}$ denotes the anticommutator. For each parameter $\theta_j$, the corresponding symmetric logarithmic derivative (SLD) $\hat{L}_j$ satisfies
\begin{equation}
    \frac{\partial\hat{\rho}(\bm{\theta})}{\partial\theta_j}
    = \frac{1}{2}\{\hat{\rho}(\bm{\theta}),\hat{L}_j\}.
    \label{eq:S2}
\end{equation}

Let $\hat{\rho}(\bm{\theta})=\sum_m p_m|e_m\rangle\langle e_m|$ be the spectral decomposition of $\hat{\rho}(\bm{\theta})$ over its support, where $p_m$ are the positive eigenvalues and $|e_m\rangle$ are the corresponding orthonormal eigenvectors. In this representation, the $(j,k)$ element of the QFIM is given by
\begin{equation}
    \mathcal{Q}_{jk}(\bm{\theta}) = \sum_m \frac{1}{p_m}\frac{\partial p_m}{\partial\theta_j}\frac{\partial p_m}{\partial\theta_k} + 4\sum_m p_m\,\mathrm{Re}\left(\frac{\partial\langle e_m|}{\partial\theta_j}\frac{\partial|e_m\rangle}{\partial\theta_k}\right) - 8\sum_{m,n}\frac{p_m p_n}{p_m+p_n}\,\mathrm{Re}\left(\frac{\partial\langle e_m|}{\partial\theta_j}|e_n\rangle\langle e_n|\frac{\partial|e_m\rangle}{\partial\theta_k}\right).
    \label{eq:S3}
\end{equation}
For a pure parameterized state $\hat{\rho}(\bm{\theta})=|\psi(\bm{\theta})\rangle\langle\psi(\bm{\theta})|$, Eq.~\eqref{eq:S3} can be simplified as
\begin{equation}
    \mathcal{Q}_{jk}(\bm{\theta})
    = 4\,\mathrm{Re}\left[
       \frac{\partial\langle\psi(\bm{\theta})|}{\partial\theta_j}
       \frac{\partial|\psi(\bm{\theta})\rangle}{\partial\theta_k}
       -
       \frac{\partial\langle\psi(\bm{\theta})|}{\partial\theta_j}
       |\psi(\bm{\theta})\rangle\langle\psi(\bm{\theta})|
       \frac{\partial|\psi(\bm{\theta})\rangle}{\partial\theta_k}
       \right].
    \label{eq:S4}
\end{equation}

Specifically, when the parameter vector $\bm{\theta}$ is encoded into a fixed pure probe state $|\psi\rangle$ by the unitary evolution $\hat{U}(\bm{\theta})$, the resulting state is $|\psi(\bm{\theta})\rangle=\hat{U}(\bm{\theta})|\psi\rangle$. The $(j,k)$ element of the QFIM can then be expressed in terms of the local generators $\hat{\mathcal{H}}_j$ and $\hat{\mathcal{H}}_k$ as
\begin{equation}
    \mathcal{Q}_{jk}(\bm{\theta}) = 4\,\mathrm{Cov}_{|\psi\rangle}\!\left(\hat{\mathcal{H}}_j,\hat{\mathcal{H}}_k\right) = \left.2\langle\{\hat{\mathcal{H}}_j,\hat{\mathcal{H}}_k\}\rangle - 4\langle\hat{\mathcal{H}}_j\rangle\langle\hat{\mathcal{H}}_k\rangle\right|_{|\psi\rangle},
    \label{eq:S5}
\end{equation}
where $\langle\hat{A}\rangle_{\psi}:=\langle\psi|\hat{A}|\psi\rangle$. The symmetrized covariance of two Hermitian operators $\hat{A}$ and $\hat{B}$ in the input state $|\psi\rangle$ is defined by
\begin{equation*}
    \mathrm{Cov}_{|\psi\rangle}\!\left(\hat{A},\hat{B}\right) := \frac{1}{2}\langle\psi|\{\hat{A},\hat{B}\}|\psi\rangle - \langle\psi|\hat{A}|\psi\rangle\langle\psi|\hat{B}|\psi\rangle.
\end{equation*}
For each parameter $\theta_j$, the associated local generator is given by
\begin{equation}
    \hat{\mathcal{H}}_j = \mathrm{i}\,\hat{U}^{\dagger}(\bm{\theta})\frac{\partial\hat{U}(\bm{\theta})}{\partial\theta_j}.
    \label{eq:S6}
\end{equation}

In DQS task, the quantity of global parameter is often a scalar function $f(\bm{\theta})$ of the distributed parameters. We introduce a local reparameterization $\bm{\xi}=(\xi_1,\ldots,\xi_M)^{\mathsf{T}}$, where $\xi_{\mathrm{I}}=\xi_1=f(\bm{\theta})$ is the parameter of interest and $\bm{\xi}_{\mathrm{N}}=(\xi_2,\ldots,\xi_M)^{\mathsf{T}}$ is the nuisance-parameter vector. The QFIM in the new coordinates is
\begin{equation}
    \mathcal{Q}(\bm{\xi}) = J_{\bm{\theta}(\bm{\xi})}^{\mathsf{T}}\mathcal{Q}\bigl(\bm{\theta}(\bm{\xi})\bigr)J_{\bm{\theta}(\bm{\xi})},
    \label{eq:S7}
\end{equation}
where $J_{\bm{\theta}(\bm{\xi})}=J_{\bm{\xi}(\bm{\theta})}^{-1}$ is the Jacobian of the inverse transformation and $[J_{\bm{\xi}(\bm{\theta})}]_{jk}=\partial\xi_j/\partial\theta_k$. Assuming that $\mathcal{Q}(\bm{\xi})$ is positive definite at the operating point, the partial QFI for $\xi_{\mathrm{I}}$ is the Schur complement of the nuisance block,
\begin{equation}
    \mathcal{Q}_{\xi_{\mathrm{I}}|\bm{\xi}_{\mathrm{N}}}(\bm{\xi}) = \mathcal{Q}_{\xi_{\mathrm{I}};\xi_{\mathrm{I}}}(\bm{\xi}) - \mathcal{Q}_{\xi_{\mathrm{I}};\bm{\xi}_{\mathrm{N}}}(\bm{\xi})\left[\mathcal{Q}_{\bm{\xi}_{\mathrm{N}};\bm{\xi}_{\mathrm{N}}}(\bm{\xi})\right]^{-1}\mathcal{Q}_{\bm{\xi}_{\mathrm{N}};\xi_{\mathrm{I}}}(\bm{\xi}).
    \label{eq:S8}
\end{equation}
where $\xi_{\mathrm{I}}=\xi_{1}$ denotes the parameter of interest and $\bm{\xi}_{\mathrm{N}}=(\xi_{2},\xi_{3},\dots,\xi_{M})^\mathsf{T}$ denotes the nuisance-parameter vector. The inverse of $\mathcal{Q}_{\xi_{\mathrm{I}}|\bm{\xi}_{\mathrm{N}}}(\bm{\xi})$ sets the achievable quantum Cramér-Rao bound (QCRB) for estimating $\xi_{\mathrm{I}}=\xi_{1}=f(\bm{\theta})$. Specifically, the variance for estimating the target scalar function $f(\bm{\theta})$ satisfies
\begin{equation}
	\mathrm{Var}\left[f(\bm{\theta})\right] = \mathrm{Var}\left(\xi_{1}\right) \ge \frac{1}{\nu\mathcal{Q}_{\xi_{\mathrm{I}}|\bm{\xi}_{\mathrm{N}}}(\bm{\xi})}, \label{eq:S9}
\end{equation}
where $\nu$ denotes the number of independent trials. 

\subsection{MILR for fixed causal-order network architecture }

We consider a free-space quantum sensing scheme in which a single probe sequentially interrogates $M$ sensors in a cyclic network. At each sensor, the local parameter $\theta$ is encoded via the unitary $\hat{U}_{\theta}=\exp(-\mathrm{i}\theta\hat{X})$. Free-space propagation over a distance $z$ between successive sensors is described by $\hat{U}_{z}=\exp(-\mathrm{i}z\hat{P}^{2}/2k)$. For a single sensing route with a fixed causal order, the total unitary evolution is given by
\begin{equation}
    \hat{U}_{FCO} = \hat{U}_{+} = \hat{U}_{z_{M}}\hat{U}_{\theta_{M}}\cdots\hat{U}_{\theta_{2}}\hat{U}_{z_{1}}\hat{U}_{\theta_{1}}\hat{U}_{z_{0}}
    \label{eq:S10}
\end{equation}
Considering the relations
\begin{equation}
	\label{eq:S11}
	\begin{split}
		\hat{U}_{z}^{\dagger}\hat{X}\hat{U}_{z} = \hat{X}+\frac{z}{k}\hat{P}, \quad \hat{U}_{z}^{\dagger}\hat{P}\hat{U}_{z} = \hat{P}, \\
		\hat{U}_{z}\hat{X}\hat{U}_{z}^{\dagger} = \hat{X}-\frac{z}{k}\hat{P}, \quad \hat{U}_{z}\hat{P}\hat{U}_{z}^{\dagger} = \hat{P},
	\end{split}
\end{equation}
and using the Baker–Campbell–Hausdorff (BCH) formula
\begin{equation*}
	\mathrm{e}^{\hat{A}}\mathrm{e}^{\hat{B}} = \mathrm{e}^{\hat{A}+\hat{B}+\frac{1}{2}[\hat{A},\hat{B}]}, 
\end{equation*}
if $\hat{A}$ and $\hat{B}$ commute with their commutator, i.e., $[\hat{A},[\hat{A},\hat{B}]]=[\hat{B},[\hat{A},\hat{B}]]=0$. We can then derive $\hat{U}^{(\mathrm{FCO})}$ as
\begin{equation}
    \begin{aligned}
        \hat{U}_{\mathrm{FCO}} &= \hat{U}_{+} = \hat{U}_{z_M}\hat{U}_{\theta_M}\cdots\hat{U}_{\theta_2}\hat{U}_{z_1}\hat{U}_{\theta_1}\hat{U}_{z_0} \\
        &= \exp\!\left[-\frac{\mathrm{i}}{2k}\sum_{j=0}^{M-1}z_j\left(\sum_{l=j+1}^{M}\theta_l\right)^2\right]\,\exp\!\left[-\mathrm{i}\frac{(M+1)\bar{z}}{2k}\hat{P}^{2}\right]\,\exp\!\left[-\frac{\mathrm{i}}{k}\sum_{j=0}^{M-1}z_j\left(\sum_{l=j+1}^{M}\theta_l\right)\hat{P}\right]\,\exp\!\left[-\mathrm{i}\sqrt{M}\,\bar\theta_0\hat{X}\right] \\
        &= \exp\!\left[-\mathrm{i}\frac{(M+1)\bar{z}}{2k}\hat{P}^{2}\right]\,\exp\!\left[-\frac{\mathrm{i}}{k}\sum_{j=0}^{M-1}z_j\left(\sum_{l=j+1}^{M}\theta_l\right)\hat{P}\right]\,\exp\!\left[-\mathrm{i}\sqrt{M}\,\bar\theta_0\hat{X}\right].
    \end{aligned}
    \label{eq:S12}
\end{equation}
where $\bar\theta_0=\frac{1}{\sqrt{M}}\sum_{j=1}^{M}\theta_j$ denotes the direction of the average of the sensing parameters, and $\bar{z}=\frac{1}{M+1}\sum_{j=0}^{M}z_{j}$ denotes the average propagation distance between adjacent nodes. The first factor is a global phase for a single route and is therefore omitted from the final probe evolution. For simplicity, we denote
\[
\Gamma_{+}(\boldsymbol{\theta})=\sum_{j=0}^{M-1} z_j\left(\sum_{l=j+1}^{M}\theta_l\right),\quad
\Gamma_{-}(\boldsymbol{\theta})=\sum_{j=1}^{M} z_j\left(\sum_{l=1}^{j}\theta_l\right),\quad
\]
Therefore
\[\Gamma_{+}(\boldsymbol{\theta})+\Gamma_{-}(\boldsymbol{\theta})=(M+1)\bar{z}\sqrt{M}\bar{\theta}_0.\]
Then the $\hat{U}^{(\mathrm{FCO})}$ can be denoted as
\begin{equation}
    \hat{U}^{(\mathrm{FCO})} = \exp\!\left[-\mathrm{i}\frac{(M+1)\bar{z}}{2k}\hat{P}^{2}\right]\,\exp\!\left[-\mathrm{i}\frac{\Gamma_{+}(\boldsymbol{\theta})}{k}\hat{P}\right]\,\exp\!\left[-\mathrm{i}\frac{\Gamma_{+}(\boldsymbol{\theta})+\Gamma_{-}(\boldsymbol{\theta})}{(M+1)\bar{z}}\hat{X}\right],
    \label{eq:S13}
\end{equation}
The probe state arriving at the server is given by $|\psi_{+}\rangle=\hat{U}_{+}|\psi_{i}\rangle$, where $|\psi_{i}\rangle$ is the initial probe state. In this single-route configuration, the target-dependent contribution $\Gamma_{+}(\boldsymbol{\theta})$ is coupled to the unknown propagation distances ${z_j}$, which therefore act as nuisance parameters. As a result, the target information encoded in $\Gamma_{+}(\boldsymbol{\theta})$ cannot be fully isolated at the server, leading to a reduction in the information available for estimating the global parameter $\bar{\theta}_{0}$. Specifically, for the sequential final state $|\psi_{+}\rangle$, the QFIM with respect to the target parameter $\bar{\theta}_{0}$ and the nuisance parameter $\Gamma_{+}(\boldsymbol{\theta})$ is given by
\begin{equation}
    \begin{split}
        \mathcal{Q}^{(\mathrm{FCO})}_{\bar{\theta}_0;\bar{\theta}_0}
        &= \frac{4(M+1)^{2}M\bar{z}^{2}(\Delta\hat{P})_{i}^{2}}{k^{2}}
        + \frac{8(M+1)M\bar{z}\operatorname{Cov}_{i}(\hat{X},\hat{P})}{k}
        + 4M(\Delta\hat{X})_{i}^{2},\\
        \mathcal{Q}^{(\mathrm{FCO})}_{\bar{\theta}_0;\Gamma_{+}(\boldsymbol{\theta})}
        &= \mathcal{Q}^{(\mathrm{FCO})}_{\Gamma_{+}(\boldsymbol{\theta});\bar{\theta}_0}
        = -\frac{4(M+1)\sqrt{M}\bar{z}(\Delta\hat{P})_{i}^{2}}{k^{2}}
        - \frac{4\sqrt{M}\operatorname{Cov}_{i}(\hat{X},\hat{P})}{k},\\
        \mathcal{Q}^{(\mathrm{FCO})}_{\Gamma_{+}(\boldsymbol{\theta});\Gamma_{+}(\boldsymbol{\theta})}
        &= \frac{4(\Delta\hat{P})_{i}^{2}}{k^{2}}, \label{eq:S14}
    \end{split}
\end{equation}
Substituting Eq.~(\ref{eq:S14}) into Eq. (\ref{eq:S8}), we obtain the partial QFI for estimating target global parameter $\bar{\theta}_{0}$ in the presence of the nuisance parameter $\Gamma_{+}(\boldsymbol{\theta})$,
\begin{equation}
\mathcal{Q}^{(\mathrm{FCO})}\left(\bar{\boldsymbol{\theta}}_{0}\middle|\hat{\rho}_{f}\right)=
\mathcal{Q}^{(\mathrm{FCO})}_{\bar{\theta}_{0}|\Gamma_{+}(\boldsymbol{\theta})}=
\mathcal{Q}^{(\mathrm{FCO})}_{\bar{\theta}_{0};\bar{\theta}_{0}}
-\mathcal{Q}^{(\mathrm{FCO})}_{\bar{\theta}_{0};\Gamma_{+}(\boldsymbol{\theta})}
\left(\mathcal{Q}^{(\mathrm{FCO})}_{\Gamma_{+}(\boldsymbol{\theta});\Gamma_{+}(\boldsymbol{\theta})}\right)^{-1}
\mathcal{Q}^{(\mathrm{FCO})}_{\Gamma_{+}(\boldsymbol{\theta});\bar{\theta}_{0}}
= 4M\left[(\Delta\hat{X})_{i}^{2}-\frac{\operatorname{Cov}_{i}(\hat{X},\hat{P})^{2}}{(\Delta\hat{P})_{i}^{2}}\right],
 \label{eq:S15}
\end{equation}

To quantify the maximum information leakage from the intermediate state at the stage $d$, we can derive $\hat{U}_d$ as
\begin{equation}
    \begin{aligned}
        \hat{U}_d
        &= \hat{U}_{\theta_d}\hat{U}_{z_{d-1}}\hat{U}_{\theta_{d-1}}
           \cdots\hat{U}_{\theta_1}\hat{U}_{z_0} \\
        &= \exp\!\left[-\frac{\mathrm{i}}{2k}
           \left(\sum_{j=0}^{d-1}z_j\right)\hat{P}^{2}\right]\,
           \exp\!\left[-\mathrm{i}\frac{\Gamma_{d}}{k}\hat{P}\right]\,
           \exp\!\left[-\mathrm{i}\sqrt{d}\,\vartheta_d\hat{X}\right],
    \end{aligned}
     \label{eq:S16}
\end{equation}
where
\[
    \Gamma_d=\sum_{j=0}^{d-1}z_j\left(\sum_{l=j+1}^{d}\theta_l\right),
    \qquad
    \vartheta_d=\frac{1}{\sqrt{d}}\sum_{l=1}^{d}\theta_l.
\]

The intermediate probe state accessible at this stage is $|\psi_d(\bm{\theta})\rangle=\hat{U}_d|\psi_i\rangle$, where $|\psi_i\rangle$ is the initial probe state. Equation~\eqref{eq:S16} shows that the dependence of the intermediate state on the sensing parameters is captured by the collective coordinates $\vartheta_d$ and $\Gamma_d$. In this two-parameter description, we take $\vartheta_d$ as the parameter of interest and treat $\Gamma_d$ as a nuisance parameter. The QFIM for $\boldsymbol{\lambda}_d=(\vartheta_d,\Gamma_d)^{\mathrm{T}}$ has elements
\begin{equation}
    \begin{aligned}
        \mathcal{Q}_{\vartheta_d;\vartheta_d}
        &= 4d\langle\Delta\hat{X}^{2}\rangle_i,
        \qquad
        \mathcal{Q}_{\vartheta_d;\Gamma_d}
        = \mathcal{Q}_{\Gamma_d;\vartheta_d}
        = \frac{4\sqrt{d}}{k}\mathrm{Cov}_i(\hat{X},\hat{P}),
        \qquad
        \mathcal{Q}_{\Gamma_d;\Gamma_d}
        = \frac{4\langle\Delta\hat{P}^{2}\rangle_i}{k^{2}}.
    \end{aligned}
    \label{eq:S17}
\end{equation}
The partial QFI for estimating $\vartheta_d$ in the presence of the nuisance parameter $\Gamma_d$ is given by the Schur complement
\begin{equation}
    \mathcal{Q}^{(\mathrm{eff})}\!\left(\vartheta_d\middle|\hat{\rho}_d\right)
    =\mathcal{Q}_{\vartheta_d\mid\Gamma_d}
    =\mathcal{Q}_{\vartheta_d;\vartheta_d}
    -\mathcal{Q}_{\vartheta_d;\Gamma_d}
     \left(\mathcal{Q}_{\Gamma_d;\Gamma_d}\right)^{-1}
     \mathcal{Q}_{\Gamma_d;\vartheta_d}.
    \label{eq:S18}
\end{equation}
At stage $d$, the normalized parameter is $\vartheta_d=\boldsymbol{\alpha}_d^{\mathrm T}\boldsymbol{\theta}/\sqrt d$, where $\boldsymbol{\alpha}_d=(\underbrace{1,\ldots,1}_{d},\underbrace{0,\ldots,0}_{M-d})^{\mathrm T}$. The global sensing direction is $\boldsymbol{\alpha}_0=\sqrt{M}(1,\ldots,1)^{\mathrm T}$, with $\bar{\theta}_0=\boldsymbol{\alpha}_0^{\mathrm T}\boldsymbol{\theta}$. For $1\leq d<M$, the component of $\boldsymbol{\alpha}_d$ orthogonal to $\boldsymbol{\alpha}_0$ is
\begin{equation}
    \boldsymbol{\alpha}_d^\perp
    =\boldsymbol{\alpha}_d
     -\left(\boldsymbol{\alpha}_0^{\mathrm T}\boldsymbol{\alpha}_d\right)\boldsymbol{\alpha}_0
    =\boldsymbol{\alpha}_d-\frac{d}{M}(1,\ldots,1)^{\mathrm T},
    \qquad
    \left\|\boldsymbol{\alpha}_d^\perp\right\|
    =\sqrt{\frac{d(M-d)}{M}}.\label{eq:S19}
\end{equation}
The unit leakage direction and its associated scalar coordinate are
\begin{equation*}
    \tilde{\boldsymbol{\alpha}}_d^\perp
    =\frac{\boldsymbol{\alpha}_d^\perp}{\left\|\boldsymbol{\alpha}_d^\perp\right\|},
    \qquad
    \tilde{\theta}_d
    =\left(\tilde{\boldsymbol{\alpha}}_d^\perp\right)^{\mathrm T}\boldsymbol{\theta}.
\end{equation*}
The collective coordinate therefore decomposes as
\begin{equation*}
    \vartheta_d
    =\sqrt{\frac{d}{M}}\,\bar{\theta}_0
     +\sqrt{\frac{M-d}{M}}\,\tilde{\theta}_d.
\end{equation*}
With $\bar{\theta}_0$ held fixed and $\Gamma_d$ treated as an independent nuisance parameter, the partial QFI along the leakage direction is
\begin{equation}
\mathcal{Q}^{(\mathrm{eff})}\!\left(\tilde{\theta}_d\middle|\hat{\rho}_d\right) = \frac{M-d}{M}\,\mathcal{Q}^{(\mathrm{eff})}\!\left(\vartheta_d\middle|\hat{\rho}_d\right) = \frac{4d(M-d)}{M}\left[\langle(\Delta\hat{X})^2\rangle_i - \frac{\mathrm{Cov}_i^2(\hat{X},\hat{P})}{\langle(\Delta\hat{P})^2\rangle_i}\right].\label{eq:S20}
\end{equation}
Combining the partial QFI for the leakage parameter $\tilde{\theta}_d$ at the intermediate node with the partial QFI for the normalized global parameter $\bar{\theta}_0$ at the server, we obtain the MILR
\begin{equation}
    \mathcal{L}
=\frac{\max_d\,\mathcal{Q}^{(\mathrm{eff})}\!\left(\tilde{\theta}_d\middle|\hat{\rho}_d\right)}
           {\mathcal{Q}\!\left(\bar{\theta}_0\middle|\hat{\rho}_f\right)}
    =\max_d\frac{d(M-d)}{M^2},\label{eq:S21}
\end{equation}
where the maximum is taken over the intermediate stages for which the two-parameter model applies. If these stages include an integer nearest to $M/2$, the maximum can be specified according to the total number of sensors in the network. For an even number of sensors, it is attained at $d=M/2$ and equals $1/4$. For an odd number of sensors, it equals $(M^2-1)/(4M^2)$. In either case, this defines a leakage level, which is independent of the generator uncertainties.

\subsection{MILR for causal-order switching network architecture}

Eq.~(\ref{eq:S21}) shows that under a fixed causal order, the MILR is independent of probe squeezing and approaches $1/4$ as the number of sensors increases. To enhance privacy, we therefore combine causal-order switching with probe squeezing between two opposite query orders in the cyclic sensing network. A control ancilla prepared in the classically mixed state $\hat{\rho}_{\mathrm{c}}^{(\mathrm{SW})}=(|0\rangle\langle 0|+|1\rangle\langle 1|)/2$ selects the two orders with equal probability. In this case, the probe passes through the cyclic network in either the forward or backward direction with equal probability. Then the final joint state of the probe and the ancilla can be derived as
\begin{equation}
	\hat{\rho}_{f}^{(\mathrm{SW})} = \frac{1}{2}\hat{U}_{+}|\psi_{i}\rangle\langle\psi_{i}|\hat{U}_{+}^{\dagger}\otimes|0\rangle\langle 0|+\frac{1}{2}\hat{U}_{-}|\psi_{i}\rangle\langle\psi_{i}|\hat{U}_{-}^{\dagger}\otimes|1\rangle\langle 1|, \label{eq:S22}
\end{equation}
where $\hat{U}_{+}$ and $\hat{U}_{-}$ are the evolution operators of forward and backward propagations given by
\begin{equation*}
	\begin{split}
		\hat{U}_{+} = \hat{U}_{z_{M}}\hat{U}_{\theta_{M}}\cdots\hat{U}_{\theta_{2}}\hat{U}_{z_{1}}\hat{U}_{\theta_{1}}\hat{U}_{z_{0}} = \exp\left[-\mathrm{i}\frac{(M+1)\bar{z}}{2k}\hat{P}^{2}\right]\exp\left(-\mathrm{i}\frac{\Gamma_{+}(\boldsymbol{\theta})}{k}\hat{P}\right)\exp\left[-\mathrm{i}\frac{\Gamma_{+}(\boldsymbol{\theta})+\Gamma_{-}(\boldsymbol{\theta})}{(M+1)\bar{z}}\hat{X}\right], \\
		\hat{U}_{-} = \hat{U}_{z_{0}}\hat{U}_{\theta_{1}}\hat{U}_{z_{1}}\hat{U}_{\theta_{2}}\cdots\hat{U}_{\theta_{M}}\hat{U}_{z_{M}} = \exp\left[-\mathrm{i}\frac{(M+1)\bar{z}}{2k}\hat{P}^{2}\right]\exp\left(-\mathrm{i}\frac{\Gamma_{-}(\boldsymbol{\theta})}{k}\hat{P}\right)\exp\left[-\mathrm{i}\frac{\Gamma_{+}(\boldsymbol{\theta})+\Gamma_{-}(\boldsymbol{\theta})}{(M+1)\bar{z}}\hat{X}\right].
	\end{split}
\end{equation*}
Here, the control ancilla serves as a classical label that encodes classical correlations, allowing the two propagation directions of the probe through the network to be distinguished. The final joint state $\hat{\rho}_{f}^{(\mathrm{SW})}$ is a mixed state with eigenvalues $p_{+}=p_{-}=1/2$ and corresponding eigenstates $|\Psi_{+}\rangle=|\psi_{+}\rangle\otimes|0\rangle$ and $|\Psi_{-}\rangle=|\psi_{-}\rangle\otimes|1\rangle$, where $|\psi_{+}\rangle=\hat{U}_{+}|\psi_{i}\rangle$ and $|\psi_{-}\rangle=\hat{U}_{-}|\psi_{i}\rangle$. Substituting this spectral decomposition into Eq.~(\ref{eq:S3}), the QFIM entries for jointly estimating $\Gamma_{+}(\boldsymbol{\theta})$ and $\Gamma_{-}(\boldsymbol{\theta})$ are
\begin{equation*}
	\mathcal{Q}_{\mu\nu}\!\left(\Gamma_{+}(\boldsymbol{\theta}),\Gamma_{-}(\boldsymbol{\theta})\middle|\hat{\rho}_{f}^{(\mathrm{SW})}\right)=2\sum_{s=\pm}\mathrm{Re}\left[\frac{\partial\langle\psi_s|}{\partial\Gamma_{\mu}(\boldsymbol{\theta})}\frac{\partial|\psi_s\rangle}{\partial\Gamma_{\nu}(\boldsymbol{\theta})}-\frac{\partial\langle\psi_s|}{\partial\Gamma_{\mu}(\boldsymbol{\theta})}|\psi_s\rangle\langle\psi_s|\frac{\partial|\psi_s\rangle}{\partial\Gamma_{\nu}(\boldsymbol{\theta})}\right],
\end{equation*}
for $\mu,\nu\in\{+,-\}$. Comparing this expression with the pure-state QFIM formula in Eq.~(\ref{eq:S4}) gives
\begin{equation*}
	\mathcal{Q}\!\left(\Gamma_{+}(\boldsymbol{\theta}),\Gamma_{-}(\boldsymbol{\theta})\middle|\hat{\rho}_{f}^{(\mathrm{SW})}\right)=\frac{1}{2}\mathcal{Q}\!\left(\Gamma_{+}(\boldsymbol{\theta}),\Gamma_{-}(\boldsymbol{\theta})\middle|\,|\psi_{+}\rangle\right)+\frac{1}{2}\mathcal{Q}\!\left(\Gamma_{+}(\boldsymbol{\theta}),\Gamma_{-}(\boldsymbol{\theta})\middle|\,|\psi_{-}\rangle\right).
\end{equation*}
The two pure state QFIMs are
\begin{equation*}
	\mathcal{Q}\!\left(\Gamma_{+}(\boldsymbol{\theta}),\Gamma_{-}(\boldsymbol{\theta})\middle|\,|\psi_{+}\rangle\right)=4\begin{pmatrix}
		\frac{(\Delta\hat{X})_i^2}{(M+1)^2\bar{z}^2}+\frac{(\Delta\hat{P})_i^2}{k^2}+\frac{2\mathrm{Cov}_i(\hat{X},\hat{P})}{(M+1)\bar{z}k}
		&
		\frac{(\Delta\hat{X})_i^2}{(M+1)^2\bar{z}^2}+\frac{\mathrm{Cov}_i(\hat{X},\hat{P})}{(M+1)\bar{z}k}\\
		\frac{(\Delta\hat{X})_i^2}{(M+1)^2\bar{z}^2}+\frac{\mathrm{Cov}_i(\hat{X},\hat{P})}{(M+1)\bar{z}k}
		&
		\frac{(\Delta\hat{X})_i^2}{(M+1)^2\bar{z}^2}
	\end{pmatrix},
\end{equation*}
\begin{equation*}
	\mathcal{Q}\!\left(\Gamma_{+}(\boldsymbol{\theta}),\Gamma_{-}(\boldsymbol{\theta})\middle|\,|\psi_{-}\rangle\right)=4\begin{pmatrix}
		\frac{(\Delta\hat{X})_i^2}{(M+1)^2\bar{z}^2}
		&
		\frac{(\Delta\hat{X})_i^2}{(M+1)^2\bar{z}^2}+\frac{\mathrm{Cov}_i(\hat{X},\hat{P})}{(M+1)\bar{z}k}\\
		\frac{(\Delta\hat{X})_i^2}{(M+1)^2\bar{z}^2}+\frac{\mathrm{Cov}_i(\hat{X},\hat{P})}{(M+1)\bar{z}k}
		&
		\frac{(\Delta\hat{X})_i^2}{(M+1)^2\bar{z}^2}+\frac{(\Delta\hat{P})_i^2}{k^2}+\frac{2\mathrm{Cov}_i(\hat{X},\hat{P})}{(M+1)\bar{z}k}
	\end{pmatrix}.
\end{equation*}
Their average yields the joint state QFIM
\begin{equation}
	\mathcal{Q}\!\left(\Gamma_{+}(\boldsymbol{\theta}),\Gamma_{-}(\boldsymbol{\theta})\middle|\hat{\rho}_{f}^{(\mathrm{SW})}\right)=4\begin{pmatrix}
		\frac{(\Delta\hat{X})_i^2}{(M+1)^2\bar{z}^2}+\frac{(\Delta\hat{P})_i^2}{2k^2}+\frac{\mathrm{Cov}_i(\hat{X},\hat{P})}{(M+1)\bar{z}k}
		&
		\frac{(\Delta\hat{X})_i^2}{(M+1)^2\bar{z}^2}+\frac{\mathrm{Cov}_i(\hat{X},\hat{P})}{(M+1)\bar{z}k}\\
		\frac{(\Delta\hat{X})_i^2}{(M+1)^2\bar{z}^2}+\frac{\mathrm{Cov}_i(\hat{X},\hat{P})}{(M+1)\bar{z}k}
		&
		\frac{(\Delta\hat{X})_i^2}{(M+1)^2\bar{z}^2}+\frac{(\Delta\hat{P})_i^2}{2k^2}+\frac{\mathrm{Cov}_i(\hat{X},\hat{P})}{(M+1)\bar{z}k}
	\end{pmatrix}.
	\label{eq:S23}
\end{equation}

To obtain the partial QFI for the normalized target parameter $\bar{\theta}_0=\sqrt{M}\sum_{j=1}^{M}\theta_j$, we apply the reparameterization rule in Eq.~(\ref{eq:S7}) to the two-parameter model. Using $\Gamma_{+}(\boldsymbol{\theta})+\Gamma_{-}(\boldsymbol{\theta})=\sqrt{M}(M+1)\bar{z}\,\bar{\theta}_0,$ the Jacobian of the inverse transformation appearing in Eq.~(\ref{eq:S7}) is
\begin{equation*}
	J_{\boldsymbol{\Gamma}(\boldsymbol{\xi})}
	=\frac{\partial\boldsymbol{\Gamma}}{\partial\boldsymbol{\xi}}
	=\begin{pmatrix}
		0&1\\
		\sqrt{M}(M+1)\bar{z}&-1
	\end{pmatrix}.
\end{equation*}
Applying Eq.~(\ref{eq:S7}) to the QFIM in Eq.~(\ref{eq:S23}) gives
\begin{equation*}
	\mathcal{Q}\!\left(\bar{\theta}_0,\Gamma_{+}(\boldsymbol{\theta})\middle|\hat{\rho}_{f}^{(\mathrm{SW})}\right)
	=J_{\boldsymbol{\Gamma}(\boldsymbol{\xi})}^{\mathsf{T}}\,
	\mathcal{Q}\!\left(\Gamma_{+}(\boldsymbol{\theta}),\Gamma_{-}(\boldsymbol{\theta})\middle|\hat{\rho}_{f}^{(\mathrm{SW})}\right)
	J_{\boldsymbol{\Gamma}(\boldsymbol{\xi})}.
\end{equation*}
The transformed QFIM is explicitly
\begin{equation*}
	\mathcal{Q}\!\left(\bar{\theta}_0,\Gamma_{+}(\boldsymbol{\theta})\middle|\hat{\rho}_{f}^{(\mathrm{SW})}\right)
	=4\begin{pmatrix}
		M(\Delta\hat{X})_i^2
		+\frac{M(M+1)^2\bar{z}^2(\Delta\hat{P})_i^2}{2k^2}
		+\frac{M(M+1)\bar{z}}{k}\mathrm{Cov}_i(\hat{X},\hat{P})
		&
		-\frac{\sqrt{M}(M+1)\bar{z}(\Delta\hat{P})_i^2}{2k^2}\\
		-\frac{\sqrt{M}(M+1)\bar{z}(\Delta\hat{P})_i^2}{2k^2}
		&
		\frac{(\Delta\hat{P})_i^2}{k^2}
	\end{pmatrix}.
\end{equation*}
Applying the Schur-complement formula in Eq.~(\ref{eq:S8}) with $\bar{\theta}_0$ as the target parameter and $\Gamma_{+}(\boldsymbol{\theta})$ as the nuisance parameter yields
\begin{equation}
	\begin{aligned}
		\mathcal{Q}\!\left(\bar{\theta}_0\middle|\hat{\rho}_{f}^{(\mathrm{SW})}\right)
		&=\mathcal{Q}^{(\mathrm{SW})}_{\bar{\theta}_0;\bar{\theta}_0}
		-\mathcal{Q}^{(\mathrm{SW})}_{\bar{\theta}_0;\Gamma_{+}(\boldsymbol{\theta})}
		\left[\mathcal{Q}^{(\mathrm{SW})}_{\Gamma_{+}(\boldsymbol{\theta});\Gamma_{+}(\boldsymbol{\theta})}\right]^{-1}
		\mathcal{Q}^{(\mathrm{SW})}_{\Gamma_{+}(\boldsymbol{\theta});\bar{\theta}_0}\\
		&=\frac{M(M+1)^2\bar{z}^2(\Delta\hat{P})_i^2}{k^2}
		+\frac{4M(M+1)\bar{z}}{k}\mathrm{Cov}_i(\hat{X},\hat{P})
		+4M(\Delta\hat{X})_i^2,
	\end{aligned}
	\label{eq:S24}
\end{equation}
where $\mathcal{Q}^{(\mathrm{SW})}_{a;b}$ denotes the corresponding entry of the transformed QFIM $\mathcal{Q}\!\left(\bar{\theta}_0,\Gamma_{+}(\boldsymbol{\theta})\middle|\hat{\rho}_{f}^{(\mathrm{SW})}\right)$.

We next evaluate the MILR for the causal-order-switching scheme. The partial QFI associated with the leakage parameter $\tilde{\theta}_d$ at each intermediate stage remains unchanged from the fixed-order scheme. Combining the forward and backward query orders enables the server to recover additional propagation-induced information about $\bar{\theta}_0$, creating an information asymmetry between the server and the intermediate stages. 

Combining Eqs.~(\ref{eq:S20}) and (\ref{eq:S24}) yields
\begin{equation}	\mathcal{L}_{\mathrm{SW}}=\frac{\max_d\,\mathcal{Q}^{(\mathrm{eff})}\!\left(\tilde{\theta}_d\middle|\hat{\rho}_d\right)}{\mathcal{Q}\!\left(\bar{\theta}_0\middle|\hat{\rho}_{f}^{(\mathrm{SW})}\right)}=\frac{4\lfloor M^2/4\rfloor}{M^2}\frac{(\Delta\hat{X})_i^2-\dfrac{\mathrm{Cov}_i^2(\hat{X},\hat{P})}{(\Delta\hat{P})_i^2}}{\dfrac{(M+1)^2\bar{z}^2(\Delta\hat{P})_i^2}{k^2}+\dfrac{4(M+1)\bar{z}\,\mathrm{Cov}_i(\hat{X},\hat{P})}{k}+4(\Delta\hat{X})_i^2}.
\label{eq:S25}
\end{equation}
Here, the maximization is taken over the same intermediate stages as in Eq.~(\ref{eq:S21}), and the second equality holds when this set includes an integer nearest to $M/2$. At the beam waist, centered HG probes have vanishing first moments, $\langle\hat{X}\rangle_i=\langle\hat{P}\rangle_i=0$, and zero position--momentum covariance, $\mathrm{Cov}_i(\hat{X},\hat{P})=0$. The MILR in Eq.~(\ref{eq:S25}) therefore simplifies to
\begin{equation}
	\mathcal{L}_{\mathrm{SW}}
	=\frac{4\lfloor M^2/4\rfloor}{M^2}
	\left[
		4+\left(\frac{(M+1)\bar{z}}{k}\right)^2
		\frac{(\Delta\hat{P})_i^2}{(\Delta\hat{X})_i^2}
	\right]^{-1}.
	\label{eq:S26}
\end{equation}
\subsection{Privacy enhancement through probe squeezing}
The dependence of $\mathcal{L}_{\mathrm{SW}}$ on the variance ratio $(\Delta\hat{P})_i^2/(\Delta\hat{X})_i^2$ motivates the use of probe squeezing to further enhance network privacy. In the harmonic-oscillator representation, the $N$th-order HG mode corresponds to the Fock state $|N\rangle$. We choose the squeezed first-order HG state $|\psi_{1,\tau}\rangle=|\tau_1\rangle=\hat{S}(\tau)|1\rangle$ as the initial probe. Its transverse wave function can be expanded in odd-order HG modes as
\begin{equation}
	\psi_1(x,\tau)
	=\frac{1}{(\cosh r)^{3/2}}
	\sum_{K=0}^{\infty}
	\frac{\sqrt{(2K+1)!}}{2^K K!}
	\left(-e^{\mathrm{i}\theta}\tanh r\right)^K
	u_{2K+1}(x),
	\label{eq:S27}
\end{equation}
where $\tau=re^{\mathrm{i}\theta}$, with squeezing strength $r=|\tau|$ and phase $\theta$. Here, $u_n(x)=\langle x|n\rangle$ denotes the normalized transverse wave function of the $n$th-order HG mode at the beam waist,
\begin{equation*}
	u_n(x)
	=\frac{1}{\sqrt{2^n n!}\,(2\pi\sigma_0^2)^{1/4}}
	H_n\!\left(\frac{x}{\sqrt{2}\sigma_0}\right)
	\exp\!\left(-\frac{x^2}{4\sigma_0^2}\right),
\end{equation*}
where $H_n$ is the $n$th-order Hermite polynomial and $\sigma_0^2$ is the transverse-position variance of the fundamental Gaussian mode. For $\theta=0$, the squeezed probe has transverse-position and transverse-momentum variances $(\Delta\hat{X})_i^2=3\sigma_0^2e^{-2r}$ and $(\Delta\hat{P})_i^2=3e^{2r}/(4\sigma_0^2)$, respectively. Substituting these variances into Eq.~(\ref{eq:S26}) gives
\begin{equation}
	\mathcal{L}_{\mathrm{SW}}^{(\tau)}
	=\frac{\lfloor M^2/4\rfloor}{M^2}
	\left[
		1+\left(\frac{(M+1)\bar{z}}{4k\sigma_0^2}\right)^2
		e^{4|\tau|}
	\right]^{-1}.
	\label{eq:S28}
\end{equation}
Thus, increasing the squeezing strength $|\tau|$ reduces the MILR and enhances network privacy.

\section{Experimental Configurations}
In this section, we provide additional experimental details on the generation of the spatially squeezed beam, the configuration of the sensing network, and the compensation for the unwanted relative phase between the polarization components.

\subsection{Generation method of spatially squeezed beam}
\begin{figure}[htp]
	\centering
	\includegraphics[width=0.65\linewidth]{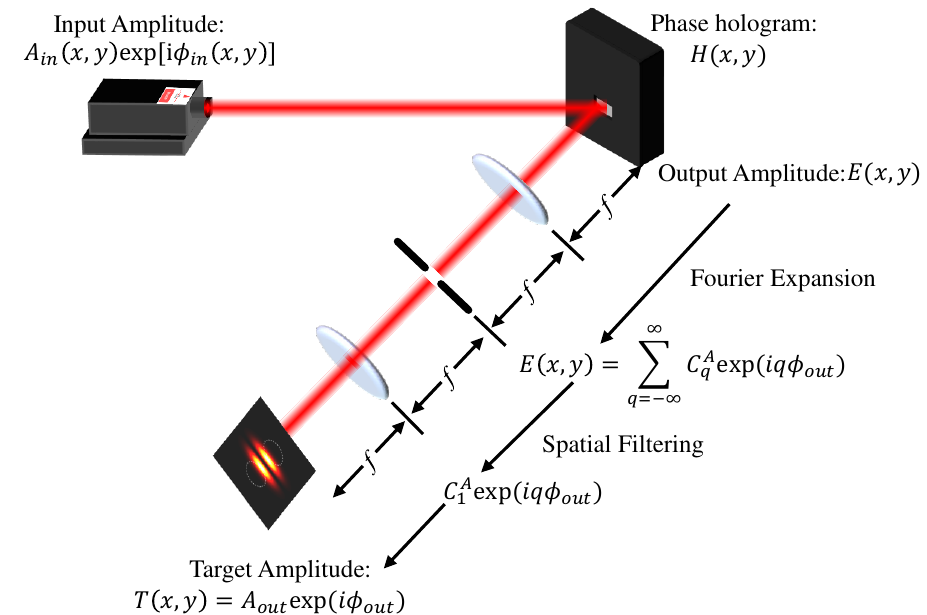}
	\caption{Generation of spatially squeezed first-order HG probes.}
	\label{fig:S1}
\end{figure}

We generate spatially squeezed first-order HG probes from an expanded Gaussian beam at $780\,\mathrm{nm}$ using a phase-only SLM and a 4-f spatial filtering system, as illustrated in Fig.~\ref{fig:S1}. The SLM displays a computer-generated hologram encoding the target wave function $\psi_1(x,\tau)$, and the spatial filter selects the $+1$ diffraction order to reconstruct the desired probe.

The target spatially squeezed first-order HG field is
\begin{equation}
	T(x,y)=\psi_1(x,\tau)u_0(y)
	=A_{\mathrm{out}}(x,y)e^{\mathrm{i}\phi_{\mathrm{out}}(x,y)},
	\label{eq:S29}
\end{equation}
where $\psi_1(x,\tau)$ is the squeezed first-order HG wave function along $x$, and $u_0(y)$ is the fundamental Gaussian mode along $y$. The incident Gaussian field $A_{\mathrm{in}}e^{\mathrm{i}\phi_{\mathrm{in}}}$ is modulated by the phase hologram $H(x,y)$ displayed on the SLM, giving $E(x,y)=A_{\mathrm{in}}e^{\mathrm{i}(\phi_{\mathrm{in}}+H)}$.

To encode the target amplitude and phase, we define $\phi_{\mathrm{r}}=\phi_{\mathrm{out}}-\phi_{\mathrm{in}}+\phi_{\mathrm{g}}$, where $\phi_{\mathrm{g}}=2\pi x/\Lambda$ is the carrier-grating phase, and the normalized relative amplitude $a=A_{\mathrm{out}}/(\mathcal{C}A_{\mathrm{in}})$, with $\mathcal{C}=\max_{x,y}(A_{\mathrm{out}}/A_{\mathrm{in}})$. The phase hologram is calculated using inverse-sinc encoding:
\begin{equation}
	H(x,y)=m(x,y)\operatorname{mod}
	\!\left[\phi_{\mathrm{r}}(x,y)-\pi m(x,y)+\pi,\,2\pi\right],
	\label{eq:S30}
\end{equation}
where $m=1-\operatorname{sinc}^{-1}(a)$ and $\operatorname{sinc}(v)=\sin(\pi v)/(\pi v)$, with the inverse taken on $0\leq v\leq1$. This encoding gives the first-order field coefficient $C_1^A=A_{\mathrm{in}}\operatorname{sinc}(1-m)=\mathcal{C}^{-1}A_{\mathrm{out}}$.

An aperture in the Fourier plane of the 4-f system selects the $+1$ diffraction order. With the carrier tilt factored out, the reconstructed field is
\begin{equation}
	E_{\mathrm{out}}(x,y)
	=C_1^A(x,y)e^{\mathrm{i}\phi_{\mathrm{out}}(x,y)}
	=\mathcal{C}^{-1}T(x,y).
\label{eq:S31}
\end{equation}
\subsection{Configuration of sensing network}

\begin{figure}[htp]
	\centering
	\includegraphics[width=0.95\linewidth]{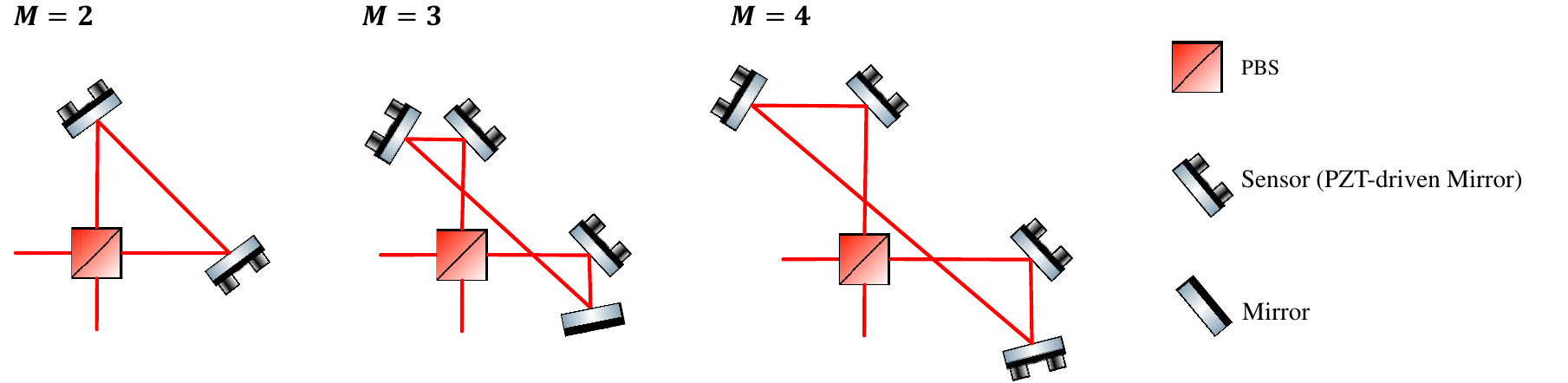}
	\caption{Configurations of the sensing network with different sensing nodes.}
	\label{fig:S2}
\end{figure}

In our experiments, we vary the number of sensing nodes by reconfiguring the polarization Sagnac interferometer, which accommodates up to four sensing nodes. The configurations for $M=2,3,4$ are shown in Fig.~\ref{fig:S2}. Each sensing node is implemented by a mirror actuated by two PZT chips. The average propagation distance between adjacent network nodes is fixed at $\bar{z}=\SI{12.5}{\centi\metre}$, giving a total cyclic propagation length of $(M+1)\bar{z}=(M+1)\times\SI{12.5}{\centi\metre}$. As sensing nodes are added, the network expands while the average node separation remains fixed.

\begin{figure}[htp]
	\centering
	\includegraphics[width=0.7\linewidth]{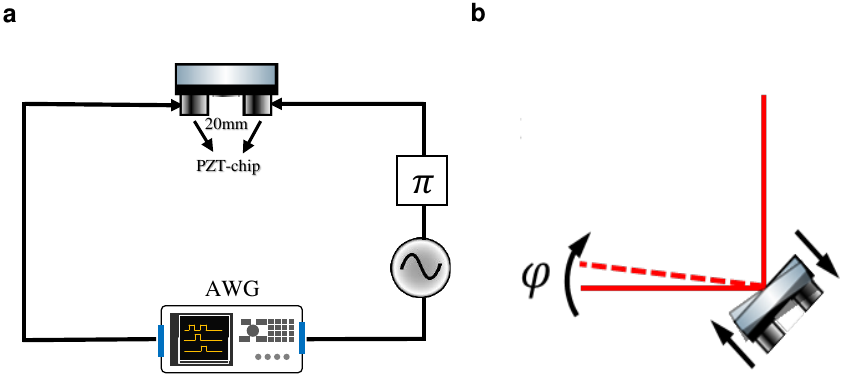}
	\caption{Implementation of the beam-tilt sensing node. \textbf{a} Two PZT chips separated by \SI{20}{\milli\metre} are driven by an arbitrary waveform generator (AWG) with a relative phase of $\pi$ to tilt the mirror. \textbf{b} The resulting angular deflection $\varphi$ of the reflected beam. }
	\label{fig:S3}
\end{figure}

Two PZT chips are bonded to the rear surface of the mirror and separated by $\SI{20}{\mm}$ along the horizontal direction. Each PZT actuator produces a displacement of approximately $\SI{22}{\nm}$ under a $\SI{1}{\V}$ drive. As shown in Figure \ref{fig:S3}\textbf{\textsf{a}}, we apply two sinusoidal signals with opposite phases to the two PZT chips. Thus, a $\SI{1}{\V}$ peak-to-peak drive signal corresponds to a $\SI{1.1}{\micro\radian}$ tilt angle on the mirror, which leads to a $\SI{2.2}{\micro\radian}$ beam-tilt modulation.
\subsection{Spatial-mode projective measurement}

\begin{figure}[htp]
	\centering
	\includegraphics[width=0.6\linewidth]{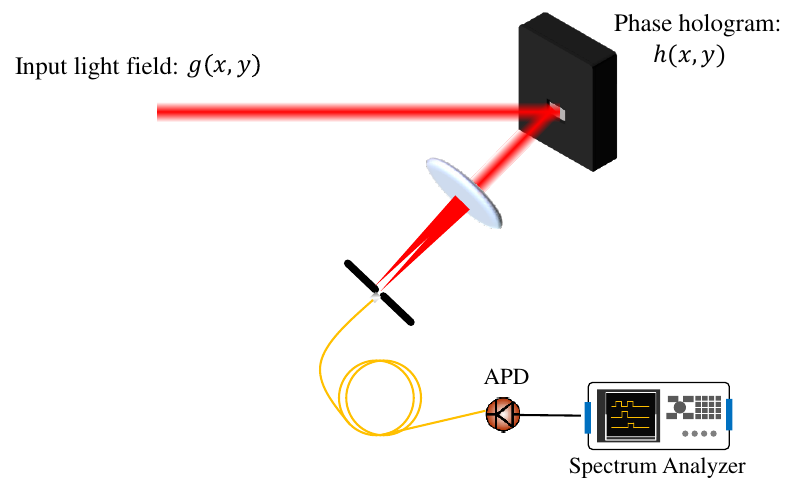}
	\caption{Schematic of spatial-mode projective measurement. The input field $g(x,y)$ is modulated by an SLM hologram encoding $h(x,y)$, Fourier-transformed by a lens, and spatially filtered by a single-mode fiber (SMF). The coupled optical power is proportional to the projection probability and is detected by an APD, whose output is analyzed using a spectrum analyzer.}
	\label{fig:S4}
\end{figure}

In our experiment, spatial-mode projective measurement is implemented using an SLM, a Fourier-transform lens, and a single-mode fiber (SMF) coupled to an APD, as shown in Figure \ref{fig:S4}. 
Let $g(x,y)$ denote the normalized transverse field incident on the SLM and $h(x,y)$ the effective complex modulation function produced by the displayed hologram. The Fourier-transform lens maps the modulated field onto the fiber input plane, yielding
\begin{equation}
E_{\mathrm{F}}(u,v)
\propto
\iint_{\mathbb{R}^{2}}
g(x,y)h(x,y)
\exp\left[-\frac{2\pi\mathrm{i}}{\lambda f}(xu+yv)\right]
\,\mathrm{d}x\,\mathrm{d}y,\label{eq:S32}
\end{equation}
where $\lambda$ is the optical wavelength, $f$ is the focal length of the lens, and $(u,v)$ are the transverse coordinates in the Fourier plane.

The SMF selects the component overlapping with its fundamental Gaussian mode, $\phi_{\mathrm{f}}(u,v)\propto
\exp[-(u^{2}+v^{2})/w_{\mathrm{f}}^{2}]$, where $w_{\mathrm{f}}$ is the fiber-mode radius. The coupling efficiency is therefore proportional to
\begin{equation}
\eta
\propto
\left|
\iint_{\mathbb{R}^{2}}
\phi_{\mathrm{f}}^{*}(u,v)
E_{\mathrm{F}}(u,v)
\,\mathrm{d}u\,\mathrm{d}v
\right|^{2}.\label{eq:S33}
\end{equation}
When the Fourier-plane field varies negligibly over the transverse extent of the fiber mode, it can be approximated by its on-axis value within the overlap integral. Consequently,
\begin{equation}
\eta
\propto
|E_{\mathrm{F}}(0,0)|^{2}
\propto
\left|
\iint_{\mathbb{R}^{2}}
g(x,y)h(x,y)
\,\mathrm{d}x\,\mathrm{d}y
\right|^{2}.\label{eq:S34}
\end{equation}
For a normalized target mode with spatial profile $m(x,y)$, encoding its complex conjugate, $h(x,y)\propto m^{*}(x,y)$, gives
\begin{equation}
\eta
\propto
\left|
\iint_{\mathbb{R}^{2}}
m^{*}(x,y)g(x,y)
\,\mathrm{d}x\,\mathrm{d}y
\right|^{2}
=
|\langle m|\psi_f(z_{\mathrm{tot}})\rangle|^{2}.\label{eq:S35}
\end{equation}
Thus, the detected optical power is proportional to the probability of projecting the incident probe onto the selected spatial mode.

For our experiment, the projection measurement $\hat{\Pi}_{1}(z_{\mathrm{tot}})=|u_{\hat{X}}^{\perp}(z_{\mathrm{tot}})\rangle\langle u_{\hat{X}}^{\perp}(z_{\mathrm{tot}})|$ is implemented by encoding the complex conjugate of the detection-mode profile on the SLM. The spatial profile $u_{\hat{X}}^{\perp}(x,z_{\mathrm{tot}})=\langle x|u_{\hat{X}}^{\perp}(z_{\mathrm{tot}})\rangle$ is
\begin{equation}
u_{\hat{X}}^{\perp}(x,z_{\mathrm{tot}}) = \frac{\mathrm{i}}{2\sqrt{6}}\sum_{K=0}^{\infty}c_{K}\Bigl[(3e^{-r}-e^{r})\sqrt{2K+2}\,u_{2K+2}(x,z_{\mathrm{tot}})-(3e^{-r}+e^{r})\sqrt{2K+1}\,u_{2K}(x,z_{\mathrm{tot}})\Bigr],\label{eq:S36}
\end{equation}
where $u_{n}(x,z_{\mathrm{tot}})=\langle x|\hat{U}_{z_{\mathrm{tot}}}|n\rangle$ denotes the $n$th-order HG mode at the detection plane, and $c_{K}$ are the HG expansion coefficients of the initial spatially squeezed probe.

\subsection{Compensation for relative phase}
\begin{figure}[htp]
	\centering
	\includegraphics[width=0.8\linewidth]{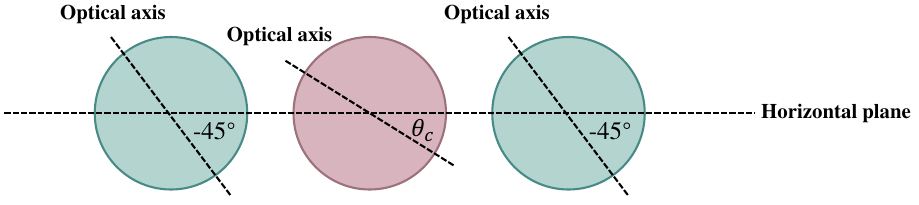}
	\caption{Schematic of polarization phase compensation. A QWP--HWP--QWP sequence compensates the relative phase offset $\delta\theta$ between the horizontal and vertical polarization components introduced by the Sagnac interferometer. Both QWPs are fixed at $-45^\circ$, and the HWP is set to $\theta_c=\delta\theta/4-\pi/4$, with all fast-axis orientations measured relative to the horizontal direction.}
	\label{fig:S5}
\end{figure}

In practice, optical imperfections in the Sagnac interferometer introduce a relative phase offset between the horizontal and vertical polarization components, modifying the polarization post-selection. We compensate this offset with a QWP--HWP--QWP sequence, as shown in Fig.~\ref{fig:S5}.
Up to a global phase, an arbitrary unitary polarization transformation admits the Euler decomposition
\begin{equation}
\hat{U}(\varphi,\xi,\zeta)
=
\hat{R}_{y}(\varphi)
\hat{R}_{z}(-\xi)
\hat{R}_{y}(\zeta),
\label{eq:S023}
\end{equation}
where $(\varphi,\xi,\zeta)$ are real parameters. We use the convention
\begin{equation*}
\hat{R}_{j}(\theta)
=
\exp(-\mathrm{i}\theta\hat{\sigma}_{j})
=
\cos\theta\,\hat{\mathbb{I}}
-\mathrm{i}\sin\theta\,\hat{\sigma}_{j},
\qquad j\in\{y,z\},
\end{equation*}
with $\hat{\sigma}_{j}$ denoting the corresponding Pauli matrix.

The Jones operator of the three-wave-plate sequence is
\begin{equation*}
\hat{U}
=
\mathrm{QWP}(\eta_{1})
\,\mathrm{HWP}(\tau)
\,\mathrm{QWP}(\eta_{2}),
\end{equation*}
where $\eta_{1}$, $\eta_{2}$, and $\tau$ denote the fast-axis orientations measured from the horizontal direction. The individual wave-plate matrices are
\begin{equation*}
\begin{aligned}
\mathrm{QWP}(\eta)
&=
\hat{R}_{y}(\eta)
\operatorname{diag}(1,\mathrm{i})
\hat{R}_{y}(-\eta),
\\
\mathrm{HWP}(\tau)
&=
\hat{R}_{y}(\tau)
\operatorname{diag}(1,-1)
\hat{R}_{y}(-\tau).
\end{aligned}
\end{equation*}
Matching this sequence to Eq.~\eqref{eq:S023} gives
\begin{equation*}
\eta_{1}=\varphi-\frac{\pi}{4},
\qquad
\eta_{2}=-\zeta-\frac{\pi}{4},
\qquad
\tau=\frac{\varphi+\xi-\zeta}{2}-\frac{\pi}{4}.
\end{equation*}

The residual phase $\delta\theta$ of the vertical component relative to the horizontal component is represented by
$\hat{U}_{\delta}=\hat{R}_{z}(\delta\theta/2)$.
The compensation operation is its inverse,
\begin{equation*}
\hat{U}_{c}
=
\hat{U}_{\delta}^{\dagger}
=
\hat{R}_{z}\left(-\frac{\delta\theta}{2}\right)
=
\operatorname{diag}
\left(
e^{\mathrm{i}\delta\theta/2},
e^{-\mathrm{i}\delta\theta/2}
\right).
\end{equation*}
Choosing $\varphi=\zeta=0$ and $\xi=\delta\theta/2$ yields the wave-plate settings
\begin{equation*}
\eta_{1}=\eta_{2}=-\frac{\pi}{4},
\qquad
\tau=\frac{\delta\theta}{4}-\frac{\pi}{4}.
\end{equation*}
These settings give $\hat{U}_{c}\hat{U}_{\delta}=\hat{\mathbb{I}}$, canceling the interferometer-induced phase offset before polarization post-selection.

\section{Supplementary Experimental Results}
For each network size $M$ and squeezing parameter $\tau$, the effective photon number $N_1$ is obtained from the measured DC optical power $I_1$ in the projection mode as $N_1=(I_1/I_0)N_0$. Here, $I_0$ is the reference optical power. The resulting values for sensing networks with $M=2,3,$ and $4$ are listed in Table~\ref{tab:S1}.
\begin{table}[hbt]
    \centering
    \begin{minipage}{\textwidth}
        \caption{Measured DC detector voltages $V_1$ and corresponding effective photon numbers $N_1$ for different network sizes $M$ and squeezing parameters $\tau$.}
        \label{tab:S1}

        \begin{tabular*}{\textwidth}{@{\extracolsep{\fill}}ccccccc@{}}
            \toprule
            & \multicolumn{2}{c}{$M=2$}
            & \multicolumn{2}{c}{$M=3$}
            & \multicolumn{2}{c}{$M=4$} \\
            \cmidrule(lr){2-3}
            \cmidrule(lr){4-5}
            \cmidrule(lr){6-7}
            $\tau$
            & $V_1$ ($\mu\mathrm{V}$) & $N_1$ ($10^{6}$)
            & $V_1$ ($\mu\mathrm{V}$) & $N_1$ ($10^{6}$)
            & $V_1$ ($\mu\mathrm{V}$) & $N_1$ ($10^{6}$) \\
            \midrule
            $0$   & $121.88$ & $1.8462$  & $95.69$  & $1.4495$  & $125.56$ & $1.9020$  \\
            $0.2$ & $191.32$ & $2.8981$  & $130.17$ & $1.9718$  & $161.31$ & $2.4435$  \\
            $0.4$ & $307.27$ & $4.6545$  & $239.16$ & $3.6228$  & $277.23$ & $4.1995$  \\
            $0.6$ & $424.09$ & $6.4241$  & $348.39$ & $5.2774$  & $354.04$ & $5.3630$  \\
            $0.8$ & $640.49$ & $9.7021$  & $567.54$ & $8.5971$  & $561.84$ & $8.5107$  \\
            $1$   & $864.65$ & $13.0977$ & $862.19$ & $13.0604$ & $823.73$ & $12.4778$ \\
            \bottomrule
        \end{tabular*}

        \par\smallskip
        {\footnotesize\raggedright
        $V_1$ denotes the DC output voltage of the APD.
        The corresponding optical power $I_1$ is obtained using
        the conversion gain of $2.65\times10^{7}\,\mathrm{V/W}$.
        The effective photon numbers are calculated as
        $N_1=(I_1/I_0)N_0=I_1\lambda T_{\mathrm{det}}/(hc)$,
        where $I_0$ is the reference optical power and
        $N_0=\dot{N}T_{\mathrm{det}}$ is the effective number of trials.
        \par}
    \end{minipage}
\end{table}

\begin{figure}[hbt]
	\centering
	\includegraphics[width=\linewidth]{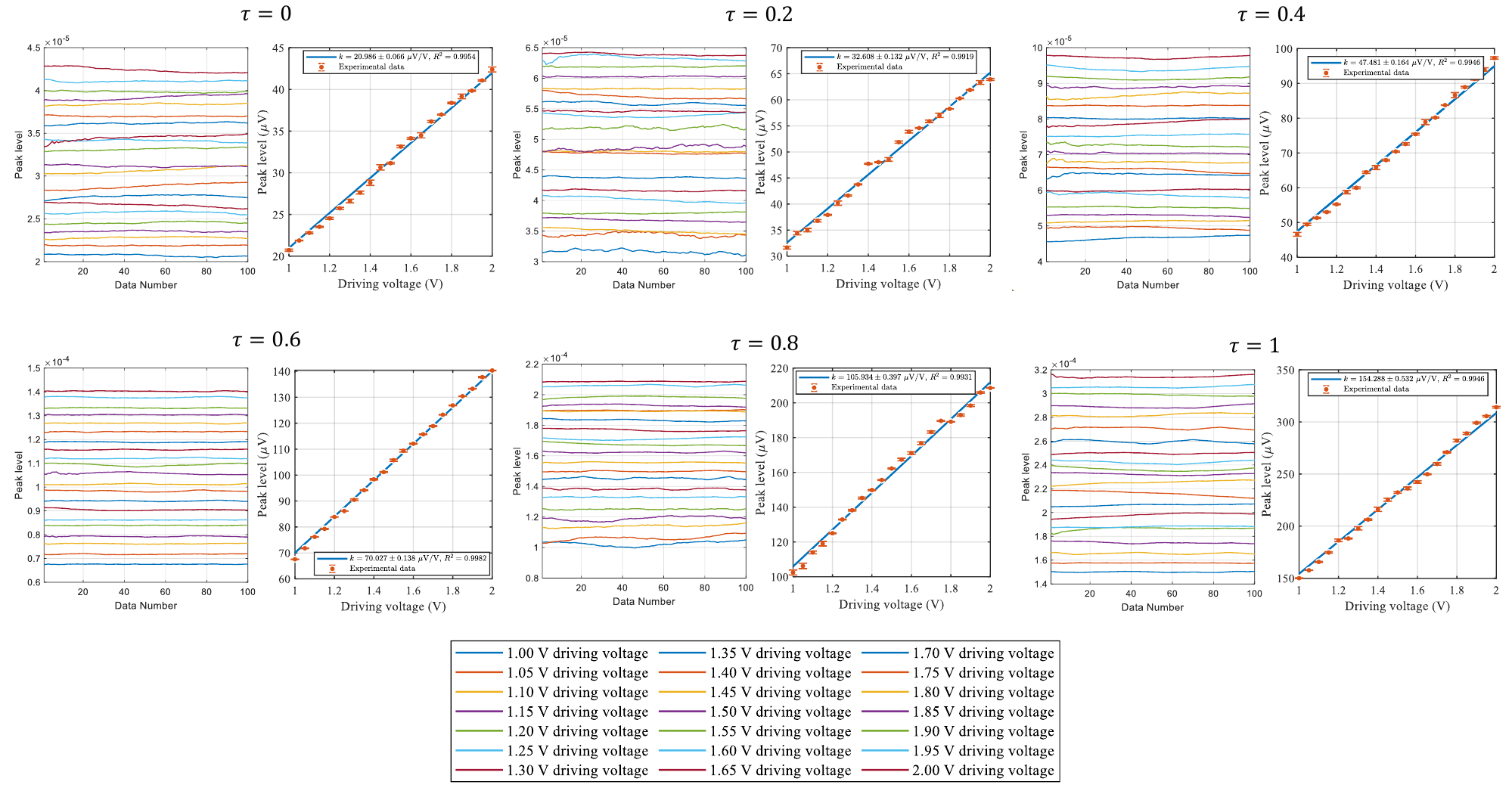}
	\caption{Measured peak levels and fitted responses for a two-sensor network ($M=2$) at different squeezing parameters $\tau$. For each $\tau$, the left panel shows the peak levels recorded by the spectrum analyzer at $2\,\mathrm{kHz}$ for peak-to-peak driving voltages from $1.00$ to $2.00\,\mathrm{V}$ in $0.05\,\mathrm{V}$ increments, ordered from bottom to top. Each trace contains 100 experimental data points. The right panel shows the mean peak level as a  linear function of driving voltage.}
	\label{fig:S6}
\end{figure}

\begin{figure}[hbt]
	\centering
	\includegraphics[width=\linewidth]{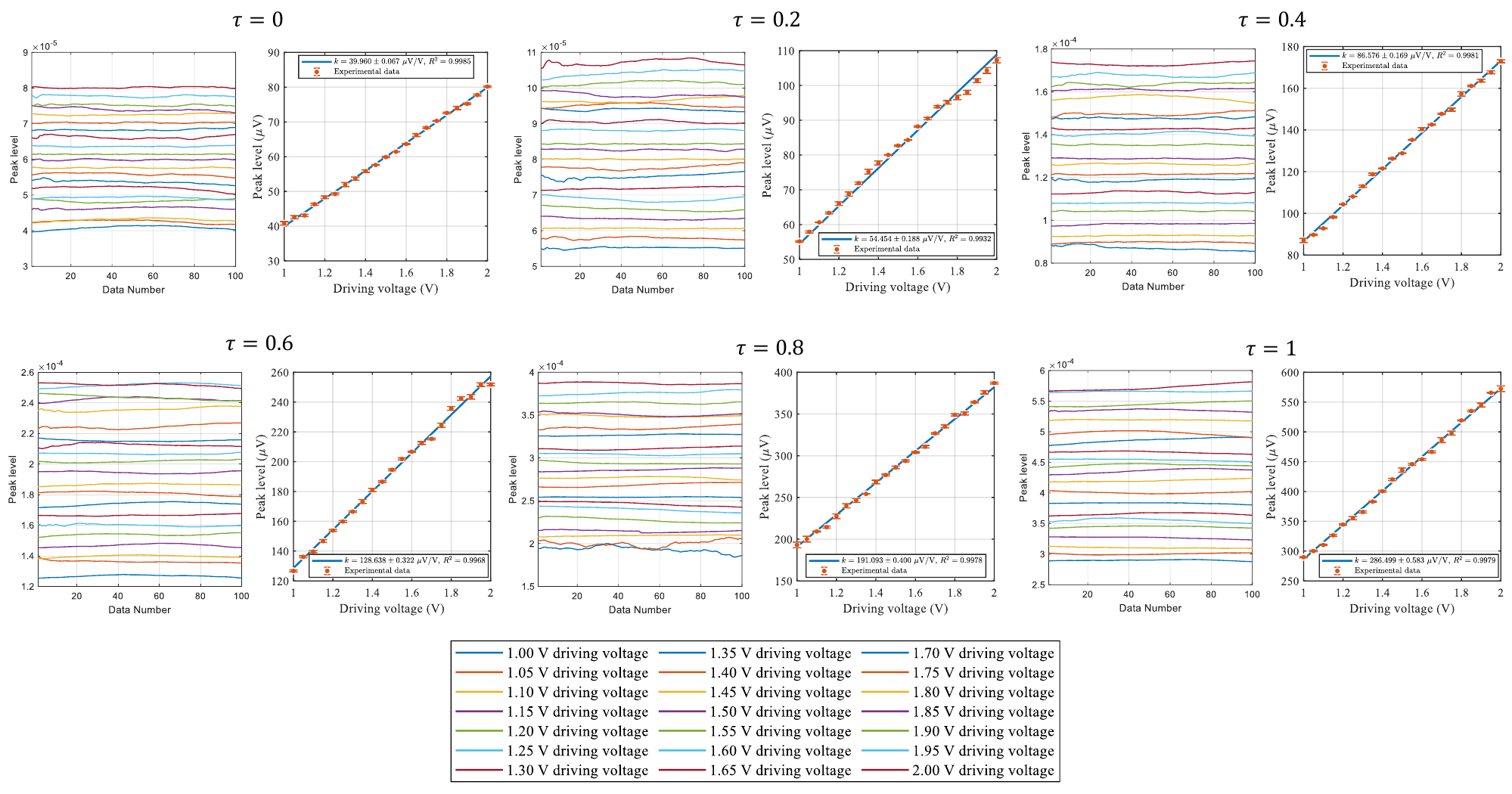}
	\caption{Measured peak levels and fitted responses for a three-sensor network ($M=3$) at different squeezing parameters $\tau$. For each $\tau$, the left panel shows the peak levels recorded by the spectrum analyzer at $2\,\mathrm{kHz}$ for peak-to-peak driving voltages from $1.00$ to $2.00\,\mathrm{V}$ in $0.05\,\mathrm{V}$ increments, ordered from bottom to top. Each trace contains 100 experimental data points. The right panel shows the mean peak level as a linear function of driving voltage.}
	\label{fig:S7}
\end{figure}

\begin{figure}[hbt]
	\centering
	\includegraphics[width=\linewidth]{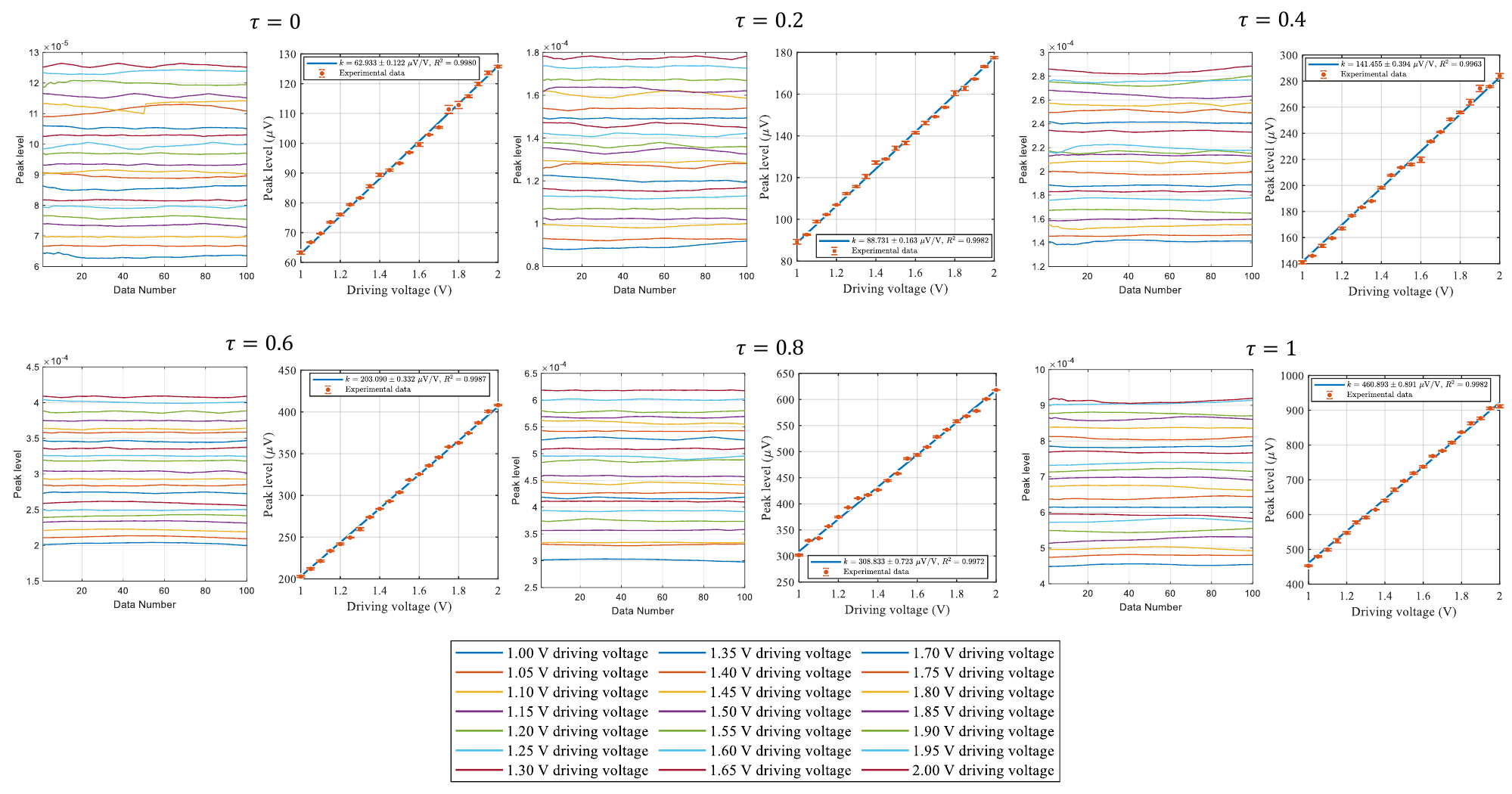}
	\caption{Measured peak levels and fitted responses for a four-sensor network ($M=4$) at different squeezing parameters $\tau$. For each $\tau$, the left panel shows the peak levels recorded by the spectrum analyzer at $2\,\mathrm{kHz}$ for peak-to-peak driving voltages from $1.00$ to $2.00\,\mathrm{V}$ in $0.05\,\mathrm{V}$ increments, ordered from bottom to top. Each trace contains 100 experimental data points. The right panel shows the mean peak level as a linear function of driving voltage. }
	\label{fig:S8}
\end{figure}

In our experiments, sinusoidal driving signals at $2\,\mathrm{kHz}$ were applied synchronously to the PZT actuators of the sensing nodes, and the corresponding peak levels were recorded using a spectrum analyzer. Measurements were performed for network sizes $M=2,3,$ and $4$ at different squeezing parameters $\tau$. For each configuration, the peak-to-peak driving voltage was varied from $1.00$ to $2.00\,\mathrm{V}$ in $0.05\,\mathrm{V}$ increments. For each network size and squeezing parameter, the experiment was repeated 100 times at each driving voltage. The mean peak level and its standard deviation were calculated from these measurements, and the mean response was fitted as a linear function of the driving voltage. The measured peak-level traces and corresponding linear fits for sensing networks with $M=2,3,$ and $4$ are presented in Figures~\ref{fig:S6}, \ref{fig:S7}, and \ref{fig:S8}, respectively.
